\documentclass[11pt,a4paper]{article}
\usepackage{xcolor,amsmath,amssymb,graphicx}
\usepackage[english]{babel}
\usepackage[T1]{fontenc}
\usepackage{ amscd,amsthm, amsfonts}
\usepackage{epsfig}
\usepackage{hhline, latexsym}
\usepackage{float}

\newlength{\dinwidth}
\newlength{\dinmargin}
\newcommand{\R}{\mathbb{R}}
\newcommand{\N}{\mathbb{N}}
\newcommand{\C}{\mathbb{C}}
\newcommand{\Z}{\mathbb{Z}}
\newcommand{\B}{\mathbb{B}}
\newcommand{\z}{\mathbf{z}}

\newcommand{\ud}{\,\mathrm{d}}
\newcommand{\Rs}{\mathcal{R}}

\newtheorem{theorem}{Theorem}[section]
\newtheorem{proposition}{Proposition}[section]

\newtheorem{remark}{Remark}[section]

\newtheorem{example}{Example}[section]

\begin{document}

\def\theequation {\thesection.\arabic{equation}}
\makeatletter\@addtoreset {equation}{section}\makeatother

\title{Breathers and solitons of \\generalized nonlinear Schrödinger equations \\as degenerations of  algebro-geometric solutions}

\author{
C.~Kalla\footnote{e-mail: Caroline.Kalla@u-bourgogne.fr;
address: Institut de Math\'ematiques de Bourgogne,
		Universit\'e de Bourgogne, 9 avenue Alain Savary, 21078 Dijon, France}}

\maketitle

\begin{abstract}
We present new solutions in terms of elementary functions of the 
multi-component nonlinear Schrödinger equations and known solutions of the Davey-Stewartson equations such as multi-soliton, breather, dromion and lump solutions. These solutions are given in a simple  determinantal form and are obtained as limiting cases in suitable degenerations of  previously derived algebro-geometric solutions.  In particular we present for the first time breather and rational breather solutions of the multi-component nonlinear Schrödinger equations.
\end{abstract}


\section{Introduction}

One of the significant advances in mathematical physics at the end of the 19th century has been the discovery by Gardner, Greene, Kruskal and Miura  \cite{GGMK} of the applicability of the Inverse Scattering Transform (IST) to the Korteweg-de Vries equation, and the construction of multi-soliton solutions. The most important physical property of solitons is that they are localized wave packets which survive collisions with other solitons without change of shape. For a guide to the vast literature on solitons, see for instance \cite{NMPZ,Des}. Existence of soliton solutions to the nonlinear Schrödinger equation (NLS)
\begin{equation}	
\mathrm{i}\,\frac{\partial \psi}{\partial t}+\frac{\partial^{2} \psi}{\partial x^{2}}+2\rho\,|\psi|^{2}\,\psi =0,  \label{NLS}
\end{equation}
where $\rho=\pm 1$, was proved by Zakharov and Shabat \cite{ZSh} using a modification of the IST.  The NLS equation is a famous nonlinear dispersive partial differential equation with 
many applications, e.g.\ in hydrodynamics (deep water waves), plasma physics
and nonlinear fiber optics.
The $N$-soliton solutions to both
the self-focusing NLS equation ($\rho=1$), as well as the defocusing NLS equation ($\rho=-1$), can also be computed by
Darboux transformations \cite{Matv}, Hirota's bilinear method
(see e.g. \cite{H,PZ,Chen}) or Wronskian techniques
(see \cite{FN,NF,F}). Hirota's method relies on a transformation of the underlying equation to a bilinear equation.  The resulting multi-soliton solutions are expressed in the form of polynomials in exponential functions.  
Wronskian techniques formulate the $N$-soliton solutions in terms
of the Wronskian determinant of $N$ functions. This method allows a straightforward direct check that the obtained solutions satisfy the equation since differentiation of a Wronskian is simple. 
On the other hand, multi-soliton solutions  of (\ref{NLS}) can be directly derived from algebro-geometric solutions when the associated hyperelliptic Riemann surface degenerates into a Riemann surface of genus zero, see for instance \cite{BBEIM}. 

In the present paper, we construct  solutions in terms of elementary functions of two generalizations of the NLS equation (\ref{NLS}): the multi-component NLS equation (n-NLS), where the number of dependent variables is increased, and the Davey-Stewartson equation (DS),  an integrable generalization to $2+1$ dimensions. 
The solutions of n-NLS and DS presented in this paper are obtained by degenerating algebro-geometric solutions, previously investigated by the author in \cite{Kalla} using Fay's identity \cite{Mum}. This method for finding solutions in terms of elementary functions has not been applied to n-NLS and DS so far.
It  provides a unified approach to various solutions of n-NLS and DS expressed in terms of a simple determinantal form, and allows to present new solutions to the multi-component NLS equation in terms of elementary functions.

One way to generalize the NLS equation is to increase the number 
of dependent variables in (\ref{NLS}). This leads to the multi-component nonlinear Schrödinger equation
\begin{equation}		
\mathrm{i}\,\frac{\partial \psi_{j}}{\partial t}+\frac{\partial^{2} \psi_{j}}{\partial x^{2}}+2\left(\sum_{k=1}^{n}s_{k}|\psi_{k}|^{2}\right)\psi_{j} =0,  \quad \quad j=1,\ldots,n,   \label{n-NLS}
\end{equation}
denoted by n-NLS$^{s}$,  where $s=(s_{1},\ldots,s_{n})$,  $s_{k}=\pm 1$. Here $\psi_{j}(x,t)$ are complex valued functions of the real variables $x$ and $t$.
The case $n=1$ corresponds to the NLS 
equation. The two-component NLS equation ($n=2$) is relevant in the
study of electromagnetic waves in optical media in which the electric field has two nontrivial components.
Integrability of the two-component NLS equation in the case $s=(1,1)$ was first established by Manakov \cite{Man}. 
In
optical fibers, for arbitrary $n\geq2$, the components $\psi_{j}$ in (\ref{n-NLS}) correspond to components of the electric field transverse to the
direction of wave propagation. These components of the transverse field form a basis of the polarization
states. Integrability for the multi-component case with any $n\geq 2$ and $s_{k}=\pm 1$ was established in \cite{RSL}. 
Multi-soliton solutions of (\ref{n-NLS}) were considered in a series of papers, see for instance \cite{Man,RL2,RL3,KLDA,APT}.

In this paper, we present a family of dark and bright multi-solitons, breather and rational breather solutions to the multi-component NLS equation. This appears to be the first time that breathers and rational breathers are given for the multi-component case. The notion of a dark soliton refers to the fact that the solution tends asymptotically to a non-zero constant, i.e., it describes a darkening on a bright background, whereas the bright soliton is a localized bright spot being described by a solution that tends asymptotically to zero. The name 'breather' reflects the behavior of the profile which is periodic in time or space and localized
in space or time. 
It is remarkable that degenerations of algebro-geometric solutions to the multi-component NLS equation lead to the breather solutions, well known in the context of the
one-component case as the soliton on a finite background
\cite{Ak et Al} (breather periodic in space), the Ma breather \cite{Ma} (breather periodic in time) and the rational
breather \cite{Per}. In the NLS framework, these solutions have been suggested as models for a
class of  extreme, freak or  rogue wave events (see e.g. \cite{HPD,OOS,AKG}). A family of rational solutions to the focusing NLS equation was constructed  in \cite{EKK} and was rediscovered recently in \cite{DM} via Wronskian techniques. Here we give for the first time a family of breather and rational breather solutions of the multi-component NLS equation.   For the one component case, our solutions consist of the well known breather and Peregrine breather of the focusing NLS equation. For the multi-component case,  we find new profiles of breathers and rational breathers which do not exist in the scalar case.

Another way to generalize the NLS equation is to increase the number of spatial dimensions to two. This leads to the DS equations,
\begin{align}		
\mathrm{i}\,\psi_{t}+\psi_{xx}-\alpha^{2}\, \psi_{yy}+2\,(\Phi+\rho\,|\psi|^{2})\,\psi &=0,     \nonumber  \\	
\Phi_{xx}+\alpha^{2}\, \Phi_{yy}+2\rho\,|\psi|^{2}_{xx} & =0,  \label{DSintro} 
\end{align}
where $\alpha=\mathrm{i},1$ and $\rho=\pm1$; $\psi(x,y,t)$ and $\Phi(x,y,t)$ are functions of the real variables $x,y$ and $t$, the latter being real valued and the former being complex valued. In what follows, DS1$^{\rho}$ corresponds to the case $\alpha=\mathrm{i}$, and DS2$^{\rho}$ to $\alpha=1$. 
The DS equation (\ref{DSintro}) was introduced in \cite{DS} to describe the evolution of a three-dimensional wave package on water of finite depth. Complete 
integrability of the equation was shown in \cite{AF}.
A main feature of equations in $1 + 1$ dimensions is the existence of
soliton solutions which are localized in one dimension. Solutions of the $2 + 1$ dimensional integrable
equations which are localized only in one dimension (plane solitons) were constructed in 
\cite{AS,APP}. Moreover, various
recurrent solutions (the growing-and-decaying mode, breather and rational growing-and-decaying
mode solutions) were investigated in \cite{TA}. 
The spectral theory of soliton type solutions to the DS1
equation (called dromions) with exponential fall off in all directions on the plane, and their connection with the initial-boundary value problem, have been
studied by different methods in a series of papers \cite{BLMP,FS,Sant,RL}. The lump solution (a rational non-singular solution) to the DS2$^{-}$ equation was discovered in \cite{APP}.

Here we present a family of dark multi-soliton solutions to the DS1 and DS2$^{+}$ equations, as well as a family of bright multi-solitons for the DS1 and DS2$^{-}$ equations, obtained by degenerating algebro-geometric solutions.
Moreover, a class of breather and rational breather solutions of the DS1 equation is given. These solutions have a very similar appearance to those in $1+1$ dimensions.
In this paper it is shown how the simplest solutions, the dromion and the lump solutions can be derived from algebro-geometric solutions.

The paper is organized as follows: Section 2 contains various facts from the theory of theta functions and identities due to Fay. These identities  were used to construct algebro-geometric solutions of n-NLS and DS equations in \cite{Kalla}, and will be needed for the degeneration of the underlying theta-functional solutions. Section 3 provides technical tools dealing with the degeneration of Riemann surfaces. We present a method which allows to degenerate algebro-geometric solutions associated to an arbitrary Riemann surface that can be applied to general integrable equations.
In Section 4  solutions in terms of elementary functions to the complexified n-NLS equation are derived by degenerating algebro-geometric solutions;  for an appropriate choice of the parameters one gets multi-solitonic solutions, and  for the first time breather and rational breather solutions to the multi-component NLS equation (\ref{n-NLS}). In Section 5 a similar program is carried out for the DS equations; well known solutions such as multi-solitons, dromion or lump are rediscovered from an algebro-geometric approach.

\section{Theta functions and Fay's identity}

Solutions of equations (\ref{n-NLS}) and (\ref{DSintro}) in terms of the multi-dimensional theta function were discussed in \cite{Kalla}. 
In this section we recall some facts from the construction of these solutions which will be used in the following  to get particular solutions as limiting cases of algebro-geometric solutions.

\subsection{Theta functions}

Let $\Rs _{g}$ be a 
compact Riemann surface of genus $g>0$. Denote by $\{\mathcal{A}_{j},\mathcal{B}_{j}\}_{j=1}^{g}$ a canonical homology basis, and by $\{\omega_{j}\}_{j=1}^{g}$ the dual basis of holomorphic differentials normalized via
\begin{equation}
\int_{\mathcal{A}_{k}}\omega_{j}=2\mathrm{i}\pi\delta_{k,\,j} \qquad k,j=1,\ldots,g. \label{norm hol diff}
\end{equation}
The matrix $\B$ of $\mathcal{B}$-periods of the normalized holomorphic differentials with entries $(\mathbb{B})_{kj}=\int_{\mathcal{B}_{k}}\omega_{j}$ 
is symmetric and has a negative definite real part. The theta function with (half integer) characteristic $\delta=[\delta', \delta'']$ is defined by
\begin{equation}
\Theta[\delta](\z|\B)=\sum_{\mathbf{m}\in\Z^{g}}\exp\left\{\tfrac{1}{2}\langle \B(\mathbf{m}+\delta'),\mathbf{m}+\delta'\rangle+\langle \mathbf{m}+\delta',\z+2\mathrm{i}\pi\delta''\rangle\right\}\label{theta}
\end{equation}
for any $\z\in\C^{g}$; here $\delta',\delta''\in \left\{0,\frac{1}{2}\right\}^{g}$ 
are the vectors of characteristic and $\langle.,.\rangle$ denotes the 
scalar product $\left\langle \mathbf{u},\mathbf{v} \right\rangle=\sum_{i}u_{i}\,v_{i}$ for any $\mathbf{u},\mathbf{v}\in\C^{g}$. The theta function $\Theta[\delta](\z)$ is even if 
the characteristic $\delta$ is even i.e, $4\left\langle 
\delta',\delta'' \right\rangle$ is even, and odd if the characteristic 
$\delta$ is odd, i.e., $4\left\langle \delta',\delta'' 
\right\rangle$ is odd. An even characteristic is called non-singular if 
$\Theta[\delta](0)\neq 0$, and an odd characteristic is called non-singular if the gradient $\nabla\Theta[\delta](0)$ is non-zero.

\subsection{Corollaries of Fay's identity} 

Let us first introduce some notation. Let $k_{a}$ denote a local parameter near $a\in\Rs_{g}$. Consider 
the following expansion of the normalized holomorphic differentials $\omega_{j}$ near $a$,
\begin{equation} 
\omega_{j}(p)= \left(V_{a,\,j}+W_{a,\,j}\,k_{a}(p)+o\left(k_{a}(p)\right)\right)\ud k_{a}(p), \label{exp hol diff}
\end{equation}
where $p$ lies in a neighbourhood of $a$, and $V_{a,\,j}, W_{a,\,j}\in\C$.
Let us denote by $D_{a}$ the operator of directional derivative along the vector $\mathbf{V}_{a}=(V_{a,1},\ldots,V_{a,g})^{t}$:
\begin{equation} 
D_{a}F(\z)=\sum_{j=1}^{g}\partial_{z_{j}}F(\z) \,V_{a,j}=\left\langle \nabla F(\z),\mathbf{V}_{a}\right\rangle, \label{2.7}
\end{equation}
where $F:\C^{g}\longrightarrow \C$ is an arbitrary function, 
and denote by $D'_{a}$ the operator of directional derivative along the vector $\mathbf{W}_{a}=(W_{a,1},\ldots,W_{a,g})^{t}$.

Now let $\delta$ be a non-singular odd 
 characteristic.  For any $\z\in\C^{g}$ and any distinct points $a,b\in\Rs_{g}$, the following two versions of Fay's identity \cite{Fay} hold (see \cite{Mum} and  \cite{Kalla})
\begin{equation} 
D_{a}D_{b}\ln\Theta(\z)\,=\,q_{1}+q_{2}\,\frac{\Theta(\z+\mathbf{r})\,\Theta(\z-\mathbf{r})}{\Theta(\z)^{2}},
\label{Fay1}
\end{equation}
\begin{equation}
D'_{a}\ln\frac{\Theta(\z+\mathbf{r})}{\Theta(\z)}+D_{a}^{2}\ln\frac{\Theta(\z+\mathbf{r})}{\Theta(\z)}+\Big(D_{a}\ln\frac{\Theta(\z+\mathbf{r})}{\Theta(\z)}-K_{1}\Big)^{2}+2\,D^{2}_{a}\ln\Theta(\z)+K_{2}=0,\label{Fay2}
\end{equation}
where  the scalars $q_{i},K_{i}$ for $i=1,2$ depend on the points $a,b$ and are given by
\begin{equation}
q_{1}(a,b)=D_{a}D_{b}\ln\Theta[\delta](\mathbf{r}), \label{q1}
\end{equation}
\begin{equation}
q_{2}(a,b)=\frac{D_{a}\,\Theta[\delta](0)\,D_{b}\,\Theta[\delta](0)}{\Theta[\delta](\mathbf{r})^{2}},\label{q2}
\end{equation}
\begin{equation}
K_{1}(a,b)=\frac{1}{2}\,\frac{D_{a}'\,\Theta[\delta](0)}{D_{a}\,\Theta[\delta](0)}+D_{a}\ln\Theta[\delta](\mathbf{r}), \label{K1}
\end{equation}
\begin{equation}
K_{2}(a,b)=-\,D'_{a}\ln\Theta(\mathbf{r})-D_{a}^{2}\ln\left(\Theta(\mathbf{r})\,\Theta(0)\right)-\Big(D_{a}\ln\Theta(\mathbf{r})-K_{1}(a,b)\Big)^{2}. \label{K2}
\end{equation}
Here we used the notation $\mathbf{r}=\int^{b}_{a}\omega$  where $\omega=(\omega_{1},\ldots,\omega_{g})^{t}$ is the vector of the normalized holomorphic differentials.

\subsection{Integral representation of $q_{2}(a,b)$ and $K_{1}(a,b)$}

Quantities $q_{2}(a,b)$ and $K_{1}(a,b)$ defined in 
(\ref{q2}) and (\ref{K1}) respectively, admit integral 
representation which will be more convenient for our purposes. These integral representations
follow from the fact that meromorphic differentials normalized by 
the condition of vanishing $\mathcal{A}$-periods can be expressed in terms of theta functions. 

Let $a,b\in\Rs_{g}$ be two distinct points connected by a contour which does not intersect $\mathcal{A}$ and $\mathcal{B}$-cycles. Hence we can define the normalized meromorphic differential of the third kind $\Omega_{b-a}$ which has residue $1$ at $b$ 
and residue $-1$ at $a$.
Now let $a\in\Rs_{g}$, and $N\in \N$ with $N>1$. The normalized
meromorphic differential of the second kind $\Omega_{a}^{(N)}$ has only one singularity at the point $a$ and is of the form 
\begin{equation}
\Omega_{a}^{(N)}(p)=\left(\frac{1}{k_{a}(p)^{N}}+O(1)\right)\mathrm{d}k_{a}(p),  \quad p\in\Rs_{g}, \label{diff 2kind}
\end{equation}
where $k_{a}$ is a local parameter in a neighbourhood of $a$.

\begin{proposition}
Let $a,b\in\Rs_{g}$ be distinct points. Denote by $k_{a}$ and $k_{b}$ local parameters in a neighbourhood of $a$ and $b$ respectively. The quantities $q_{2}(a,b)$ and $K_{1}(a,b)$ defined in 
(\ref{q2}) and (\ref{K1}) respectively admit the following integral representations:
\begin{equation}
q_{2}(a,b)=-\lim_{\begin{smallmatrix}
\tilde{b}\rightarrow b \\
\tilde{a}\rightarrow a
\end{smallmatrix}}
\left[\left(k_{a}(\tilde{a})\,k_{b}(\tilde{b})\right)^{-1}\exp\left\{\int_{\tilde{a}}^{\tilde{b}}\Omega_{b-a}(p)\right\}\right], \label{q2 int}
\end{equation}
where the integration contour does not 
cross any cycle of canonical basis, and 
\begin{equation}
K_{1}(a,b)=\lim_{
\tilde{a}\rightarrow a
}\Bigg[\int^{\tilde{a}}_{c}\Omega_{a}^{(2)}(p)+\frac{1}{k_{a}(\tilde{a})}\Bigg]-\int^{b}_{c}\Omega_{a}^{(2)}(p),\label{K1 int} 
\end{equation}
where $c$ is an arbitrary point on $\Rs_{g}$.
\end{proposition}

\noindent
Proof of (\ref{q2 int}) can be found in \cite{Kalla}, where similar statements lead to (\ref{K1 int}).

\section{Uniformization map and degenerate Riemann surfaces}

It is well known that solutions in terms of theta functions are almost periodic due to the periodicity properties of the theta functions. In the limit when the Riemann surface degenerates to a surface of genus zero, periods of the  surface diverge, and the theta series breaks down to elementary functions. 
Whereas this procedure is well-known in the case of a hyperelliptic surface, i.e., a two-sheeted branched covering of the Riemann sphere,  where such a degeneration consists in colliding branch points pairwise, it has not been applied so far to theta-functional solutions on non-hyperelliptic surfaces.

We present here a method to treat this case based on the uniformization theorem for Riemann surfaces.
In particular, we show that the theta function tends to a finite sum of exponentials in the limit when  the arithmetic genus 
of the associated Riemann surface drops to zero, and give explicitly the constants (\ref{q1})-(\ref{K2}) in this limit. As illustrated in Section 4 and 5,  particular solutions of n-NLS and DS such as multi-solitons, well known in the theory of soliton equations, arise from such degenerations of algebro-geometric solutions.

\subsection{Degeneration to genus zero}

Let us first recall some techniques  used for degenerating Riemann surfaces (see \cite{Fay} for more details).  
There exist basically two ways for
degenerating a Riemann surface by pinching a cycle: a cycle homologous to
zero in the first case, and a cycle non-homologous to zero in the second
case. The first degeneration leads to two Riemann surfaces whose genera add up to the genus of the pinched surface, whereas the limiting situation for the second degeneration is one Riemann surface of genus $g-1$ with two points identified, $g$ being the genus of the non-degenerated surface.
In both cases, locally one can identify the pinched region to a hyperboloid 
\begin{equation}
y^{2}=x^{2}-\epsilon,  \label{hyperboloid}
\end{equation}
where $\epsilon>0$ is a small parameter, such that the vanishing cycle coincides with the homology class of a closed contour around the cut  $[-\sqrt{\epsilon},\sqrt{\epsilon}]$ in the $x$-plane.
In what follows, we deal with the degeneration of the second type and make consecutive pinches until the surface degenerates to genus zero.

To degenerate the Riemann surface $\mathcal{R}_{g}$ of genus $g$ into a 
Riemann surface  $\mathcal{R}_{0}$ of genus zero, we pinch all 
$\mathcal{A}_{i}$-cycles into double points. After desingularization 
one gets  $\mathcal{R}_{0}$, and each
double point corresponds to two different points on 
$\mathcal{R}_{0}$, denoted by $u_{i}$ and $v_{i}$ for $i=1,\ldots 
,g$. In this limit, holomorphic normalized differentials $\omega_{i}$ 
become normalized differentials of the  third kind with poles at 
$u_{i}$ and $v_{i}$. Note that the
normalized differential of the second kind $\Omega_{a}^{(N)}$ with a 
pole of order $N> 1$ at $a$ remains a differential of the second 
kind with the same order of the pole after degeneration to genus zero. We 
keep the same notation for the differential of the second kind on the degenerated surface.

The compact Riemann surface $\mathcal{R}_{0}$ of genus zero is conformally equivalent to the Riemann sphere with the coordinate $w$. This mapping between $\mathcal{R}_{0}$ and the $w$-sphere is called the  uniformization map and we denote it by $w(p)=w$ for any $p\in\Rs_{0}$. Therefore, in what follows we let $\mathcal{R}_{0}$ stand also for the Riemann sphere with the coordinate $w$. 

Meromorphic differentials on $\mathcal{R}_{0}$ can be constructed using the fact that in genus zero, such differentials are entirely defined by their behaviors near their singularities. This leads to the following third and second kind differentials on $\mathcal{R}_{0}$:
\\\\
$\bullet$ \textit{Differentials of the third kind:}\\
\begin{equation}
\Omega_{v_{i}-u_{i}}=\left(\frac{1}{w-w_{v_{i}}}-\frac{1}{w-w_{u_{i}}}\right) \ud w. \label{diff3}
\end{equation}
\\
$\bullet$ \textit{Differentials of the second kind:}\\
\begin{equation}
\Omega_{a}^{(2)}=\frac{1}{k_{a}'(w_{a})}\,\frac{\ud w}{(w-w_{a})^{2}} , \label{diff2}
\end{equation}
where $k_{a}$ is a local parameter in a neighbourhood of $w_{a}\in\mathcal{R}_{0}$ and the prime denotes the derivative with respect to the argument.
This is the differential on $\Rs_{0}$, obtained from $\Omega_{a}^{(2)}$ (\ref{diff 2kind}) defined on $\Rs_{g}$, in the limit as the surface $\Rs_{g}$ degenerates to $\Rs_{0}$. The factor $(k_{a}'(w_{a}))^{-1}$ ensures that the biresidue of $\Omega_{a}^{(2)}$ with respect to the local parameter $k_{a}$ is $1$ as before the degeneration.

\subsection{Degenerate theta function}

To study the theta function with zero characteristic in the limit when the genus tends to zero, let us first analyse the behavior of the matrix $\mathbb{B}$ of $\mathcal{B}$-periods of the normalized holomorphic differentials.
Since holomorphic normalized differentials $\omega_{i}$ 
become differentials of the third kind with poles at 
$u_{i}$ and $v_{i}$, for a small parameter $\epsilon>0$, 
elements $(\mathbb{B})_{ik}$ of the matrix $\mathbb{B}$ have the following behavior
\begin{align}
(\mathbb{B})_{ik}&=\int_{u_{i}}^{v_{i}}\Omega_{v_{k}-u_{k}} +O(\epsilon), \qquad i\neq k, \label{lim B}\\
(\mathbb{B})_{kk}&=\ln \epsilon +O(1). \nonumber
\end{align}
Therefore, the real parts of diagonal terms of the Riemann matrix tend to $-\infty$ when $\epsilon$ tends to zero, that is when the Riemann surface degenerates into the Riemann surface $\Rs_{0}$. It follows that the theta function (\ref{theta}) with zero 
characteristic tends to one, since only the term corresponding to the vector $\mathbf{m}=0$ in the series may give a non-zero contribution.

To get non constant solutions of (\ref{n-NLS}) and (\ref{DSintro}) after the degeneration of the Riemann surface, let us write the argument of the theta-function in the form $\mathbf{Z}-\mathbf{D}$, where $\mathbf{D}$ is a vector with components  $D_{k}=(1/2)\,(\mathbb{B})_{kk}+d_{k}$, for some $d_{k}\in\C$ independent of $\epsilon$. Hence for any $\mathbf{Z}\in\C^{g}$ one gets
\begin{equation}
\lim_{\epsilon\rightarrow 0}\Theta(\mathbf{Z}-\mathbf{D})= \sum_{\mathbf{m}\in\left\{0,1\right\}^{g}}\exp\left\{\sum_{1\leq i<k\leq g}(\mathbb{B})_{ik}\,m_{i}m_{k}+\sum_{k=1}^{g} m_{k}\,(Z_{k}-d_{k})\right\}. \label{lim theta}
\end{equation}
Here we use the same notation for the quantities $(\mathbb{B})_{ik}$ on the degenerated surface.
The expression in the right hand side of (\ref{lim theta}) can be put  into a determinantal form (see Proposition \ref{det theta}) which will be used in the whole paper. This determinantal form can be obtained from the following representation of the components $(\mathbb{B})_{ik}$ after degeneration, obtained from (\ref{diff3}) and (\ref{lim B}),
\begin{equation}
(\mathbb{B})_{ik}=\ln\left\{\frac{w_{v_{i}}-w_{v_{k}}}{w_{v_{i}}-w_{u_{k}}}\,\frac{w_{u_{i}}-w_{u_{k}}}{w_{u_{i}}-w_{v_{k}}}\right\}.\label{B}\end{equation}
Hence, following \cite{Mat} one gets

\begin{proposition} For any $\z\in\C^{g}$ the following holds
\begin{equation}
\sum_{\mathbf{m}\in\left\{0,1\right\}^{g}}\exp\left\{\sum_{1\leq i<k\leq g}(\mathbb{B})_{ik}\,m_{i}m_{k}+\sum_{k=1}^{g} m_{k}\,z_{k}\right\}=\det(\mathbb{T}), \label{T}
\end{equation}
where $\mathbb{T}$ is a $g\times g$ matrix with entries
\begin{equation}
(\mathbb{T})_{ik}=\delta_{i,k}+\frac{w_{v_{i}}-w_{u_{i}}}{w_{v_{i}}-w_{u_{k}}}\,e^{\frac{1}{2}(z_{i}+z_{k})}. \label{element T}
\end{equation} \label{det theta}
\end{proposition}

\subsection{Degenerate constants}
The next step is to give explicitly the quantities (independent of the vector $\z$)
 appearing in (\ref{Fay1}) and (\ref{Fay2}), i.e.,
$\textbf{V}_{a},\textbf{W}_{a},\textbf{r},q_{2},$ etc,  after the degeneration to genus zero. We use the same notation for these quantities on the degenerated surface. For any distinct points $a,b\in\Rs_{0}$,  it follows from (\ref{exp hol diff}) and (\ref{diff3})  that
\begin{align}
V_{a,k}&=\frac{1}{k'_{a}(w_{a})}\,\left(\frac{1}{w_{a}-w_{v_{k}}}-\frac{1}{w_{a}-w_{u_{k}}}\right),\label{V}\\ \nonumber\\
W_{a,k}&=\frac{1}{k'_{a}(w_{a})^{2}}\,\left(-\,\frac{1}{(w_{a}-w_{v_{k}})^{2}}+\frac{1}{(w_{a}-w_{u_{k}})^{2}}\right)-\frac{k_{a}''(w_{a})}{k_{a}'(w_{a})^{2}}\,V_{a,k}\,, \label{W}\\
\nonumber\\
r_{k}&=\ln\left\{\frac{w_{b}-w_{v_{k}}}{w_{b}-w_{u_{k}}}\,\frac{w_{a}-w_{u_{k}}}{w_{a}-w_{v_{k}}}\right\},\label{r}
\end{align}
for  $k=1,\ldots, g$.
Moreover, from the integral representation of $q_{2}(a,b)$ and $K_{1}(a,b)$ (see  (\ref{q2 int}) and (\ref{K1 int})), using (\ref{diff3}) and (\ref{diff2}) one gets
\begin{align}
q_{2}(a,b)&=\frac{1}{k'(w_{a})k'(w_{b})(w_{a}-w_{b})^{2}},\label{q2 deg}\\
K_{1}(a,b)&=\frac{1}{k'(w_{a})(w_{b}-w_{a})}-\frac{1}{2}\frac{k''(w_{a})}{k'\,^{2}(w_{a})}. \label{K1 deg}
\end{align}
Putting $\z=0$ in (\ref{Fay1}) and taking the limit $\epsilon\rightarrow 0$ leads to
\begin{equation}
q_{1}(a,b)=-\,q_{2}(a,b), \label{q1 deg}
\end{equation}
due to the fact that the theta function tends to one and that its partial 
derivatives tend to zero. In the same way,
taking  the limit $\epsilon\rightarrow 0$ in (\ref{Fay2}) one gets
\begin{equation}
K_{2}(a,b)=-\,\Big(K_{1}(a,b)\Big)^{2}. \label{K2 deg}
\end{equation}

\section{Degenerate algebro-geometric solutions of n-NLS}

One way to construct solutions of (\ref{n-NLS}) is first to solve its complexified version,  a system of $2n$ equations of $2n$ dependent variables $\left\{\psi_{j},\psi_{j}^{*}\right\}_{j=1}^{n}$,
\begin{align}		
\mathrm{i}\,\frac{\partial \psi_{j}}{\partial t}+\frac{\partial^{2} \psi_{j}}{\partial x^{2}}+2\left(\sum_{k=1}^{n}\,\psi_{k}\,\psi_{k}^{*}\right)\psi_{j} &=0  ,  \nonumber  \\ 
-\mathrm{i}\,\frac{\partial \psi_{j}^{*}}{\partial t}+\frac{\partial^{2} \psi_{j}^{*}}{\partial x^{2}}+2\left(\sum_{k=1}^{n}\,\psi_{k}\,\psi_{k}^{*}\right)\psi_{j}^{*} &=0 , \quad \quad j=1,\ldots,n  ,\label{ass n-NLS} 
\end{align}
where $\psi_{j}(x,t)$ and $\psi_{j}^{*}(x,t)$ are complex valued functions of the real variables $x$ and $t$. This system reduces to the n-NLS$^{s}$ equation (\ref{n-NLS}) under the \textit{reality conditions }
\begin{equation}
\psi_{j}^{*}=s_{j}\,\overline{\psi_{j}}, \quad\quad j=1,\ldots, n. \label{real cond n-NLS}
\end{equation}
Algebro-geometric solutions of the system (\ref{ass n-NLS}) were
obtained in \cite{Kalla} by the use of the degenerated versions (\ref{Fay1}) and (\ref{Fay2}) of Fay's identity; these solutions are given by:

\begin{theorem}
Let $\mathcal{R}_{g}$ be a compact Riemann surface of genus $g>0$ and let $f$ be a meromorphic function
of degree $n+1$ on $\Rs_{g}$. Let $z_{a}\in\C$ be a non critical value of $f$, and consider the fiber $f^{-1}(z_{a})=\left\{a_{1},\ldots,a_{n+1}\right\}$ over $z_{a}$. Choose the local parameters $k_{a_{j}}(p)=f(p)-z_{a}$, 
for any point $p\in\Rs_{g}$ lying in a neighbourhood of $a_{j}$.
Let $\mathbf{D}\in\C^{g}$ and $A_{j}\neq 0$ be arbitrary constants. Then the following functions  $\left\{\psi_{j}\right\}_{j=1}^{n}$ and  $\left\{\psi_{j}^{*}\right\}_{j=1}^{n}$ are solutions of the system (\ref{ass n-NLS})
\begin{align}    
\psi_{j}(x,t)&=A_{j}\,\frac{\Theta(\mathbf{Z}-\mathbf{D}+\mathbf{r}_{j})}{\Theta(\mathbf{Z}-\mathbf{D})}\,\exp\left\{ \mathrm{i}\,(-E_{j}\,x+\,F_{j}\,t)\right\},\nonumber\\ 
\psi^{*}_{j}(x,t)&=\frac{q_{2}(a_{n+1},a_{j})}{A_{j}}\,\frac{\Theta(\mathbf{Z}-\mathbf{D}-\mathbf{r}_{j})}{\Theta(\mathbf{Z}-\mathbf{D})}\,\exp\left\{ \mathrm{i}\,(E_{j}\,x-\,F_{j}\,t)\right\}. \label{sol n-NLS comp}
\end{align}
Here $\Theta$ denotes the theta function (\ref{theta}) with zero characteristic, and $\mathbf{Z}=\mathrm{i}\,\mathbf{V}_{a_{n+1}}\,x+\,\mathrm{i}\,\mathbf{W}_{a_{n+1}}\,t,$
where vectors $\mathbf{V}_{a_{n+1}}$ and $\mathbf{W}_{a_{n+1}}$ are defined in (\ref{exp hol diff}). Moreover, $\mathbf{r}_{j}=\int^{a_{j}}_{a_{n+1}}\omega$, where $\omega$ is the vector of normalized holomorphic differentials, and
the scalars $E_{j},F_{j}$ are given by
\begin{equation}
E_{j}=K_{1}(a_{n+1},a_{j}),\qquad F_{j}=K_{2}(a_{n+1},a_{j})-2\sum_{k=1}^{n}q_{1}(a_{n+1},a_{k}). \label{E,N n-NLS}
\end{equation}
The scalars $q_{i},K_{i}$ for $i=1,2$ are defined in (\ref{q1})-(\ref{K2}). 
\end{theorem}

\noindent
The proof of this theorem is based on the following identity:
\begin{equation}
\sum_{k=1}^{n+1}\mathbf{V}_{a_{k}}=0, \label{sum V}
\end{equation}
which is satisfied by the vectors $\mathbf{V}_{a_{k}}$ associated to the fiber $f^{-1}(z_{a})=\left\{a_{1},\ldots,a_{n+1}\right\}$ over $z_{a}$.
We shall use this relation to construct solutions of (\ref{n-NLS}) in terms of elementary functions.

\begin{remark}
\rm{The relationship between solutions of the Kadomtesv-Petviashvili (KP1) equation (generalization of the KdV equation to two spatial variables, see, for instance, \cite{BBEIM}) and solutions of the multi-component NLS equation was investigated in \cite{Kalla}. This relationship implies that 
 all solutions of equation (\ref{n-NLS}) constructed in this paper provide also solutions of the KP1 equation as explained in \cite{Kalla}.}
\end{remark}

In the next section,  solutions of (\ref{ass n-NLS}) in terms of elementary functions are derived from solutions (\ref{sol n-NLS comp}) by degenerating the associated Riemann surface $\Rs_{g}$ into a Riemann surface of genus zero.
Imposing reality conditions (\ref{real cond n-NLS}), by an appropriate choice of the parameters one gets special solutions of (\ref{n-NLS}) such  as multi-solitons and breathers.
To  the best of our knowledge, such an approach to multi-solitonic solutions of n-NLS$^{s}$ has not been studied before. Moreover, breather and rational breather solutions to the multi-component case are derived here for the first time.

\subsection{Determinantal solutions of the complexified n-NLS equation}

Solutions of the complexified scalar NLS equation in terms of elementary functions were obtained in \cite{BBEIM}, when the genus of the associated hyperelliptic spectral curve tends to zero. For specific choices of parameters, they get dark and bright multi-solitons of the NLS equation, as well as quasi-periodic modulations of the plane wave solutions previously constructed in \cite{IRS}. A direct generalization of this approach to the multi-component case is not obvious, due to the complexity of the associated spectral curve. To bypass this problem and to construct spectral data associated to algebro-geometric solutions (\ref{sol n-NLS comp}) in the limit when the genus tends to zero, we use the uniformization map between the degenerate Riemann surface and the sphere. Details of such a degeneration were presented in Section 3.

Let us discuss solutions of n-NLS in genus zero. Consider the following meromorphic function $f(w)$ on the sphere: 
\begin{equation}
f(w)=\alpha\prod_{i=1}^{n+1}\frac{w-w_{a_{i}}}{w-w_{b_{i}}} \label{fct f}
\end{equation}
where $w_{a_{j}}\neq w_{b_{k}}$ for all $j,k$, $w_{a_{j}}\neq w_{a_{k}}$ for $j\neq k$, and $\alpha\in\C$. Without loss of generality, put $\alpha=1$. This function is of 
degree $n+1$ on the sphere, hence it represents a genus zero (n+1)-sheeted 
branched covering of $\C \mathbb{P}^{1}$. Recall that a meromorphic function $f$ on the sphere is called real if its zeros as well as its poles are real or pairwise conjugate.

If not stated otherwise, the local parameter in a neighbourhood of a regular point $w_{a}$ (i.e. $f'(w_{a})\neq 0$) is chosen to be $k_{a}(w)=f(w)-f(w_{a})$ for any $w$ lying in a neighbourhood of $w_{a}$.
Solutions of the complexified system (\ref{ass n-NLS}) associated to the meromorphic function $f$ (\ref{fct f}) on the sphere are given by:

\begin{proposition} Let $j,k\in\N$ satisfy $1\leq j\leq n$ and $1\leq k\leq g$. Let $f$ be a meromorphic function (\ref{fct f}) of degree $n+1$ on the sphere, with complex zeros $\{w_{a_{i}}\}_{i=1}^{n+1}$ and complex poles  $\{w_{b_{i}}\}_{i=1}^{n+1}$. Let $\mathbf{d}\in\C^{g}$ and $A_{j}\neq 0$ be arbitrary constants. Moreover, assume that $w_{u_{k}},w_{v_{k}}\in\C$ satisfy
\begin{equation}
f(w_{u_{k}})=f(w_{v_{k}}). \label{cond n-NLS}
\end{equation}
Then the following functions are solutions of the complexified system (\ref{ass n-NLS})
\begin{align}
\psi_{j}(x,t)&=A_{j}\,\frac{\det(\mathbb{T}_{j,1})}{\det(\mathbb{T}_{j,0})}\,\exp \{\mathrm{i}\,(-E_{j}\,x+F_{j}\,t)\}, \nonumber\\
\psi_{j}^{*}(x,t)&= \frac{q_{2}(a_{n+1},a_{j})}{A_{j}}\,\frac{\det(\mathbb{T}_{j,-1})}{\det(\mathbb{T}_{j,0})}\,\exp \{\mathrm{i}\,(E_{j}\,x-F_{j}\,t)\}.  \label{deg sol n-NLS} 
\end{align}
For $\beta=-1,0,1$, $\mathbb{T}_{j,\beta}$ denotes the $g\times g$ matrix with entries (\ref{element T})
where $z^{j}_{k}=Z_{k}-d_{k}+\beta\,r_{j,k}$. Here $Z_{k}=\mathrm{i}\,V_{a_{n+1},k}\,x+\mathrm{i}\,W_{a_{n+1},k}\,t,$
where the scalars $V_{a_{n+1},k}$ and $W_{a_{n+1},k}$ are defined in (\ref{V}) and (\ref{W}), and
$r_{j,k}$ is defined in (\ref{r}) with $w_{a}:=w_{a_{n+1}}$ and $w_{b}:=w_{a_{j}}$.
The scalars $E_{j}$ and $F_{j}$ are given by
\[E_{j}=K_{1}(a_{n+1},a_{j}),\qquad F_{j}=-\,(K_{1}(a_{n+1},a_{j}))^{2}+2\sum_{k=1}^{n}q_{2}(a_{n+1},a_{k}), \]
where $q_{2}(a_{n+1},a_{j})$ and $K_{1}(a_{n+1},a_{j})$ are defined in (\ref{q2 deg}) and (\ref{K1 deg}). \label{prop deg sol n-NLS}
\end{proposition}

\begin{proof}
Consider solutions (\ref{sol n-NLS comp}) associated to a Riemann surface $\Rs_{g}$ of genus $g$, and assume $f(a_{i})=0$ for any $1\leq i\leq n+1$. Pinch all $\mathcal{A}$-cycles of the associated Riemann surface $\Rs_{g}$ into double points, as explained in Section 3. After desingularization, the meromorphic function $f$ of degree $n+1$ on $\Rs_{g}$ becomes a meromorphic function of degree $n+1$ on the sphere, given in general form by (\ref{fct f}).
In the limit considered here, the theta function tends to the determinantal form (\ref{T}). Quantities defined on the degenerated surface and independent of the variables $x$ and $t$ were constructed in Section 3.3 and are given in (\ref{V})-(\ref{K2 deg}). 
Condition (\ref{cond n-NLS}) follows from the fact that double points appearing after degeneration of $\Rs_{g}$ are desingularized into two distinct points $w_{u_{k}}$ and $w_{v_{k}}$ having the same projection under the meromorphic function $f$.
Note that equation (\ref{sum V}) holds in the limit, since by (\ref{exp hol diff}) and (\ref{fct f}) one has
\begin{equation}
\sum_{i=1}^{n+1}V_{a_{i},k}=\frac{1}{f(w_{u_{k}})}-\frac{1}{f(w_{v_{k}})} \label{deg sum V}
\end{equation}
which by (\ref{cond n-NLS}) equals zero for $k=1,\ldots,g$.
\end{proof}

\begin{remark}
\rm{Functions (\ref{deg sol n-NLS}) give a family of  solutions to the complexified multi-component NLS equation (\ref{ass n-NLS}) depending on $3n+g+2$ complex parameters: 
$w_{a_{i}},w_{b_{i}}$ for $1\leq i\leq n+1$, $d_{k}$ for $1\leq k\leq g$, and $A_{j}$ for $1\leq j\leq n$.} 
\end{remark}

\begin{remark}
\rm{The following transformations leave  equation (\ref{ass n-NLS}) invariant
\begin{align}
\psi_{j}(x,t)&\longrightarrow\psi_{j}\left(\beta\,x+2\beta\lambda\,t,\beta^{2}\,t\right)\,\exp\left\{ -\mathrm{i}\left(\lambda\,x+\lambda^{2}\,t\right)\right\},\nonumber\\
\psi^{*}_{j}(x,t)&\longrightarrow\beta^{2}\,\psi^{*}_{j}\left(\beta\,x+2\beta\lambda\,t,\beta^{2}\,t\right)\,\exp\left\{ \mathrm{i}\left(\lambda\,x+\lambda^{2}\,t\right)\right\}, 
\label{trans sol n-NLS}
\end{align}
where $\lambda=\mu\,\beta^{-1}$ for any $\mu\in\C$ and any $\beta\neq 0$. Such a transformation may be useful to
simplify the expressions in the obtained solutions and thus to facilitate
the numerical implementation. }
\end{remark}

\subsection{Multi-solitonic solutions of n-NLS}

Imposing reality conditions (\ref{real cond n-NLS}) on the degenerate solutions (\ref{deg sol n-NLS}) of the complexified system, 
one gets particular solutions of (\ref{n-NLS}) such as dark and bright multi-solitons. Dark and bright solitons differ by 
the fact that the modulus of the first tends to a non zero 
constant and the modulus of the second tends to zero when the spatial variable tends to infinity.
Such solutions were obtained in \cite{BBEIM} for the one component case by degenerating algebro-geometric solutions, and describe elastic collisions between solitons. Elastic means that
the solitons asymptotically retain their shape and speed after interaction. The interaction of
vector solitons is more complex than the one of scalar solitons because inelastic collisions can appear in all components of one solution (see for instance \cite{APT}). 
\\\\
In what follows $N\in\N$ with $N\geq 1$.

\subsubsection{Dark multi-solitons of n-NLS$^{s}$, $s\neq(1,\ldots,1)$.} Dark multi-soliton solutions of 
2-NLS$^{s}$ were investigated in \cite{RL2}. The dark $N$-soliton solution derived here corresponds to elastic interactions between $N$ dark solitons. Moreover, it is shown that this type of solutions does not exist for the focusing multi-component nonlinear Schrödinger equation, i.e., in the case where $s=(1,\ldots,1)$.

\begin{proposition}  Let $j,k\in\N$ satisfy $1\leq j\leq n$ and $1\leq k\leq N$. Let $f$ be a real meromorphic function (\ref{fct f}) of degree $n+1$ on the sphere, having $n+1$ real zeros $\{w_{a_{i}}\}_{i=1}^{n+1}$. Choose $\theta\in\R$ and $\mathbf{d}\in\R^{N}$. Moreover, assume that $w_{u_{k}},w_{v_{k}}\in\C$ satisfy (\ref{cond n-NLS}) and
\begin{equation}
\overline{w_{u_{k}}}=w_{v_{k}}. \label{dark uv}
\end{equation}
Put $s_{j}=\text{sign}(f'(w_{a_{n+1}})f'(w_{a_{j}}))$. Then the following functions define smooth dark $N$-soliton solutions of n-NLS$^{s}$, where $s=(s_{1},\ldots,s_{n})$ with $s\neq(1,\ldots,1)$,
\begin{equation}
\psi_{j}(x,t)=A_{j}\,e^{\mathrm{i}\theta}\,\frac{\det(\mathbb{T}_{j,1})}{\det(\mathbb{T}_{j,0})}\,\exp \left\{\mathrm{i}\,(-E_{j}\,x+F_{j}\,t)\right\}.
\label{dark n-NLS} 
\end{equation}
Here $A_{j}=|q_{2}(a_{n+1},a_{j})|^{1/2},$ and the remaining notation is as in Proposition \ref{prop deg sol n-NLS} with $g=N$.   \label{prop dark NLS}
\end{proposition}

\begin{proof}
Let us check that the functions $\psi_{j}$ and $\psi_{j}^{*}$ defined in (\ref{deg sol n-NLS}) satisfy reality conditions (\ref{real cond n-NLS}) with $s_{j}=\text{sign}(f'(w_{a_{n+1}})f'(w_{a_{j}}))$.
Put $A_{j}=|q_{2}(a_{n+1},a_{j})|^{1/2}$ in (\ref{deg sol n-NLS}). Then with the above  assumptions, it is straightforward to see that $\psi_{j}^{*}=s_{j}\,\overline{\psi_{j}}$ where 
$s_{j}=\text{sign}(q_{2}(a_{n+1},a_{j}))$, which by (\ref{q2 deg}) leads to $s_{j}=\text{sign}(f'(w_{a_{n+1}})f'(w_{a_{j}}))$. Moreover, with (\ref{dark uv}) it can be seen that condition (\ref{sum V}) is equivalent to
\begin{equation}
\frac{1}{|w_{a_{n+1}}-w_{v_{k}}|^{2}}+\sum_{j=1}^{n}\frac{f'(w_{a_{n+1}})}{f'(w_{a_{j}})}\,\frac{1}{|w_{a_{j}}-w_{v_{k}}|^{2}}=0, \label{no foc}
\end{equation}
for $k=1,\ldots,N$. Therefore, by (\ref{no foc}) the quantity $f'(w_{a_{n+1}})f'(w_{a_{j}})$ cannot be positive for all $j$, which yields $s\neq(1,\ldots,1)$. 
The solutions are smooth since the denominator in (\ref{dark n-NLS}) is a finite sum of real exponentials.
\end{proof}

\begin{remark}
\rm{The dark $N$-soliton solutions (\ref{dark n-NLS}) depend on $N+1$ real parameters $d_{k},\theta$ and a real meromorphic function $f$ (\ref{fct f}) defined by $2n+2$ real parameters. The solitons are dark since  the modulus of the $\psi_{j}$ tends to $A_{j}$ when $x\in\R$ tends to infinity.
}
\end{remark}

\begin{example} 
With the notation of Proposition \ref{prop deg sol n-NLS} and \ref{prop dark NLS}, functions $\psi_{j}$ (\ref{dark n-NLS}) are given for $N=1$ by 
\[\psi_{j}(x,t)=A_{j}\,\frac{1+e^{Z_{1}-d_{1}+r_{j,1}}}{1+e^{Z_{1}-d_{1}}}\,e^{\mathrm{i}\,(-E_{j}x+F_{j}t)}.\]
\end{example}

\subsubsection{Bright multi-solitons of n-NLS$^{s}$.}

Bright multi-solitons of the NLS equation presented in \cite{BBEIM} were obtained by collapsing all branch cuts of
the underlying hyperelliptic curve of the algebro-geometric solutions. This way they get solutions expressed as the  quotient of a finite sum of exponentials similar to dark multi-solitons, except that the modulus of the solutions tends to zero instead of a non-zero constant when the spatial variable tends to infinity.
Following this approach, a family of bright multi-solitons of 
n-NLS$^{s}$ is obtained here by further degeneration of (\ref{deg sol n-NLS}).

For the multi-component case there exist two sorts of bright soliton interactions: elastic or inelastic. Inelastic collisions between bright solitons were investigated in \cite{RL3} for the two component case and in \cite{KLDA} for the multi-component case. 
The family of bright multi-solitons of 
n-NLS$^{s}$ obtained here describes the standard elastic collision with phase shift. Notice that there exist various ways to degenerate algebro-geometric solutions. Therefore, it appears possible that bright solitons with inelastic collision can be obtained by different degenerations.

\begin{proposition} Let $j\in\N$ satisfy $1\leq j\leq n$. Take $w_{a_{j}},\theta\in\R$ and choose $\mathbf{\hat{d}}\in\C^{2N}$ such that $\overline{\hat{d}_{2k-1}}=\hat{d}_{2k}$. Moreover, let $w_{u_{2k}},w_{v_{2k-1}}\in\C$ satisfy
\begin{equation}
 \overline{w_{u_{2k}}}=w_{v_{2k-1}}  \label{bright uv}
\end{equation}
for $1\leq k\leq N$. Choose $\gamma_{j}\in\R$ and put $s_{j}=\text{sign}(\gamma_{j})$. Then the following functions give bright $N$-soliton solutions of n-NLS$^{s}$ 
\begin{equation}
\psi_{j}(x,t)=A_{j}\,e^{\mathrm{i}\theta}\,\frac{\det(\mathbb{K}_{j})}{\det(\mathbb{M})}, 
\label{bright n-NLS} 
\end{equation}
where $A_{j}=|\gamma_{j}|^{1/2}\,|w_{a_{j}}|^{-1}$. Here
$\mathbb{K}_{j}$ and $\mathbb{M}$ are $2N\times2N$ matrices with entries $(\mathbb{K}_{j})_{ik}$ and $(\mathbb{M})_{ik}$ given by:
\begin{eqnarray}
\text{- for $i$ and $k$ even:}& (\mathbb{K}_{j})_{ik}&\hspace{-0.2cm}=\delta_{2,i}\,\frac{w_{u_{i}}}{w_{u_{k}}}\,e^{\frac{1}{2}(z_{2}+z_{k}+\hat{r}_{j,2}+\hat{r}_{j,k})}+\delta_{i,k}-\delta_{2,i}\,\delta_{2,k}\nonumber\\
&&\,\,+\,\delta_{2,k}\,(\delta_{2,i}-1)\,\frac{w_{u_{i}}}{w_{u_{2}}}\,e^{\frac{1}{2}(z_{i}-z_{2}+\hat{r}_{j,i}-\hat{r}_{j,2})} \nonumber\\
\text{- for $i$ even and $k$ odd:}& (\mathbb{K}_{j})_{ik}&\hspace{-0.2cm}=\alpha_{u_{k}}^{2}\,\frac{w_{u_{i}}}{\alpha_{v_{i}}-\alpha_{u_{k}}}\,\frac{\alpha_{v_{2}}-\alpha_{v_{i}}}{\alpha_{v_{2}}-\alpha_{u_{k}}}\,e^{\frac{1}{2}(z_{i}+z_{k}+\hat{r}_{j,i}+\hat{r}_{j,k})}\nonumber\\
\text{- for $i$ odd and $k$ even:}& (\mathbb{K}_{j})_{ik}&\hspace{-0.2cm}=\frac{w_{v_{i}}}{w_{v_{i}}-w_{u_{k}}}\,e^{\frac{1}{2}(z_{i}+z_{k}+\hat{r}_{j,i}+\hat{r}_{j,k})}
 \nonumber\\
\text{- for $i$ and $k$ odd:}& (\mathbb{K}_{j})_{ik}&\hspace{-0.2cm}=\delta_{i,k} \nonumber\\\nonumber
\\
\text{- for $i,k$ even, or $i,k$ odd:}&(\mathbb{M})_{ik}&\hspace{-0.2cm}=\delta_{i,k}\nonumber\\ 
\text{- for $i$ even and $k$ odd:}&(\mathbb{M})_{ik}&\hspace{-0.2cm}=\alpha_{u_{k}}\,\alpha_{v_{i}}\,\frac{w_{u_{i}}}{\alpha_{v_{i}}-\alpha_{u_{k}}}\,e^{\frac{1}{2}(z_{i}+z_{k})} \nonumber\\
\text{- for $i$ odd and $k$ even:}&(\mathbb{M})_{ik}&\hspace{-0.2cm}=-\,\frac{w_{v_{i}}}{w_{v_{i}}-w_{u_{k}}}\,e^{\frac{1}{2}(z_{i}+z_{k})}.
 \nonumber
\end{eqnarray}
Here $\overline{\alpha_{v_{2k}}}=\alpha_{u_{2k-1}}$ where 
\begin{equation}
\alpha_{u_{2k-1}}=\sum_{j=1}^{n}\gamma_{j}\left(\frac{1}{w_{a_{j}}}-\frac{1}{w_{a_{j}}-w_{v_{2k-1}}}\right). \label{bright alpha}
\end{equation}
Moreover, $z_{k}$ is a linear function of the variables $x$ and $t$ satisfying $\overline{z_{2k}}=z_{2k-1}$, given by
\begin{equation}
z_{2k-1}=\mathrm{i}\,\alpha_{u_{2k-1}}\,x+\mathrm{i}\,\alpha_{u_{2k-1}}^{2}\,t-\hat{d}_{2k-1}.\nonumber
\end{equation}
The scalars $\hat{r}_{j,k}$ satisfy $\overline{\hat{r}_{j,2k}}=-\,\hat{r}_{j,2k-1}$  where
\begin{equation}
\hat{r}_{j,2k-1}=\ln\left\{\frac{w_{a_{j}}-w_{v_{2k-1}}}{w_{a_{j}}\,w_{v_{2k-1}}\,\alpha_{u_{2k-1}}}\right\}. \nonumber
\end{equation}  \label{prop bright NLS}
\end{proposition}

\begin{proof}
Consider functions (\ref{deg sol n-NLS}) obtained from (\ref{sol n-NLS comp}) for the choice of local parameters $k_{a_{i}}$:
\[k_{a_{i}}(w)=(\gamma_{i}\,f'(w_{a_{i}}))^{-1}f(w)\]
 for any $w$ lying in a neighbourhood of $w_{a_{i}}$, $i=1,\ldots,n+1$, and assume $g=2N$. Hence condition (\ref{sum V}) becomes
\begin{equation}
 \sum_{i=1}^{n+1}\gamma_{i}\left(\frac{1}{w_{a_{i}}-w_{v_{k}}}-\frac{1}{w_{a_{i}}-w_{u_{k}}}\right)=0 \label{sum V bright}
\end{equation}
for $k=1,\ldots,N$.
Now put $A_{j}=|q_{2}(a_{n+1},a_{j})|^{1/2}$ in (\ref{deg sol n-NLS}), where \[q_{2}(a_{n+1},a_{j})=\gamma_{n+1}\,\gamma_{j}\,(w_{a_{n+1}}-w_{a_{j}})^{-2}.\]
Choose a small parameter $\epsilon>0$ and define
$ d_{k}=-\ln \epsilon+\hat{d}_{k}, $
for  $k=1,\ldots,2N,$ and 
\begin{equation}
w_{u_{2k-1}}=w_{a_{n+1}}+\epsilon^{2}\,\alpha_{u_{2k-1}}^{-1}, \quad w_{v_{2k}}=w_{a_{n+1}}+\epsilon^{2}\,\alpha_{v_{2k}}^{-1},
\end{equation}
for $k=1,\ldots,N$.
Now put $\gamma_{n+1}=\epsilon^{2}$ and consider in the determinant $\det(\mathbb{T}_{j,1})$ appearing in (\ref{deg sol n-NLS}) the substitution 
\[L_{2i}\longrightarrow L_{2i}-\frac{(\mathbb{T}_{j,1})_{2i,2}}{(\mathbb{T}_{j,1})_{2,2}}\,L_{2},\]
for $i=2,\ldots,N,$ where $L_{k}$ denotes the line number $k$ of the matrix $\mathbb{T}_{j,1}$, and $(\mathbb{T}_{j,1})_{i,k}$ denotes the entries of this matrix. In the limit $\epsilon\rightarrow 0$, it can be seen that the functions $\psi_{j}$ given in (\ref{deg sol n-NLS}) converge towards functions (\ref{bright n-NLS}), where the following change of parameters (eliminating the parameter $w_{a_{n+1}}$) has been made:
\[w_{a_{j}}\rightarrow w_{a_{j}}+w_{a_{n+1}},\qquad w_{u_{2k}}\rightarrow w_{u_{2k}}+w_{a_{n+1}}, \qquad w_{v_{2k-1}}\rightarrow w_{v_{2k-1}}+w_{a_{n+1}},\]
for $j=1,\ldots,n$ and $k=1,\ldots,N$.
 Analogous statements can be made for the functions $\psi_{j}^{*}$. 
By assumption, it is straightforward to see that the functions $\psi_{j}$ and $\psi_{j}^{*}$ obtained in the limit considered here satisfy the reality conditions 
$\psi_{j}^{*}=s_{j}\,\overline{\psi_{j}}$ with
$s_{j}=\text{sign}(\gamma_{j})$. Moreover, in this limit condition (\ref{sum V bright}) yields (\ref{bright alpha}).
\end{proof}

\begin{remark}
\rm{The bright $N$-soliton solutions (\ref{bright n-NLS}) depend on $2N$ complex parameters $\hat{d}_{2k-1}$, $w_{v_{2k-1}}$, and $2n+1$ real parameters $w_{a_{j}},\gamma_{j},\theta$.
Moreover, all parameters appearing in (\ref{bright n-NLS}) are free, contrary to the dark multi-solitons (\ref{dark n-NLS}) where parameters $w_{u_{k}}$ and $w_{v_{k}}$ have to satisfy the polynomial equation (\ref{cond n-NLS}). The solitons are bright since the modulus of the $\psi_{j}$ tends to zero when $x\in\R$ tends to infinity, in contrast to the dark solitons.}
\end{remark}


\subsection{Breather and rational breather solutions of n-NLS}

Solutions obtained here differ from the dark multi-solitons studied in Section 4.2.1 by the reality condition imposed on parameters $w_{u_{i}}$ and $w_{v_{i}}$ in solutions (\ref{deg sol n-NLS}) of the complexified system for $i=1,\ldots,g$.  By an appropriate choice of parameters, one gets periodic solutions (breathers) as well as rational solutions (rational breathers). 
The name `breather' reflects the behavior of the profile of the solution which is periodic in time (respectively, space) and localized
in space (respectively, time).
This appears to be the first time that explicit breather and rational breather solutions of n-NLS$^{s}$ are given.
\\\\
In what follows $N\in\N$ with $N\geq 1$.

\subsubsection{Multi-Breathers of n-NLS$^{s}$.} 
Multi-breather solutions of n-NLS$^{s}$ are given in the following 
proposition. The $N$-breather solution corresponds to an elastic interaction between $N$ breathers.

\begin{proposition} Let $j,k\in\N$ satisfy $1\leq j\leq n$ and $1\leq k\leq N$.  Let $f$ be a real meromorphic function (\ref{fct f}) of degree $n+1$ on the sphere, having $n+1$ real zeros $\{w_{a_{i}}\}_{i=1}^{n+1}$. Choose $\theta\in\R$ and take $\mathbf{\hat{d}}\in\C^{2N}$ such that $\overline{\hat{d}_{2k-1}}=\hat{d}_{2k}$. Let $w_{u_{2k}},w_{u_{2k-1}},w_{v_{2k}},w_{v_{2k-1}}\in\C$ satisfy (\ref{cond n-NLS}) and 
\begin{equation}
\overline{w_{u_{2k}}}=w_{v_{2k-1}}, \quad \overline{w_{u_{2k-1}}}=w_{v_{2k}}.  \label{uv breather}
\end{equation}
Put $s_{j}=\text{sign}(f'(w_{a_{n+1}})f'(w_{a_{j}}))$. Then the following functions define $N$-breather solutions of n-NLS$^{s}$
\begin{equation}
\psi_{j}(x,t)=A_{j}\,e^{\mathrm{i}\theta}\,\frac{\det(\mathbb{T}_{j,1})}{\det(\mathbb{T}_{j,0})}\,\exp \left\{\mathrm{i}\,(-E_{j}\,x+F_{j}\,t)\right\},
\label{breather n-NLS} 
\end{equation}
where $A_{j}=|q_{2}(a_{n+1},a_{j})|^{1/2},$ and the remaining notation is the same as in Proposition \ref{prop deg sol n-NLS} for $g=2N$. 
\end{proposition}

\begin{remark}
\rm{Functions (\ref{breather n-NLS}) cover a family of breather solutions of n-NLS$^{s}$ depending on $N$ complex parameters $d_{k}$, a real parameter $\theta$, and a real meromorphic function $f$ (\ref{fct f}) defined by $2n+2$ real parameters. }
\end{remark}

To simplify the computation of the solutions, we apply transformation (\ref{trans sol n-NLS}) to the solutions (\ref{breather n-NLS}), with $\beta$ and $\lambda$ given by
\begin{equation}
\beta=1, \qquad \lambda=\frac{1}{2}\,f''(w_{a_{n+1}})\,f'(w_{a_{n+1}})^{-2}. \label{tranf breather}
\end{equation} 
Hence, the quantity $f''(w_{a_{n+1}})f'(w_{a_{n+1}})^{-2}\,V_{a_{n+1},k}$ in the expression (\ref{W}) for the scalar $W_{a_{n+1},k}$ disappears, as well as the quantity $\frac{1}{2}f''(w_{a_{n+1}})f'(w_{a_{n+1}})^{-2}$ in the expression (\ref{K1 deg}) for the scalar $K_{1}(a_{n+1},a_{j})$.

\begin{figure}[hbtp]  
\begin{center}
\includegraphics[width=130mm,height=90mm]{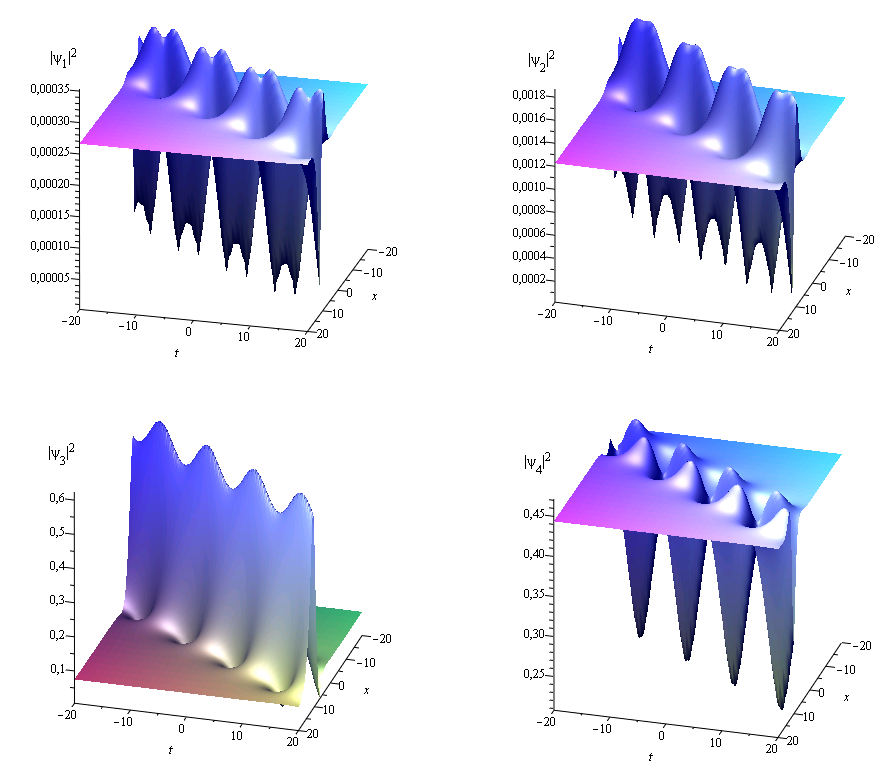}  
\end{center}
\caption{Breather of $4$-NLS$^{--+-}$. \label{breather4NLS}}
\end{figure}

\begin{figure}[hbtp]  
\begin{center}
\includegraphics[width=130mm,height=90mm]{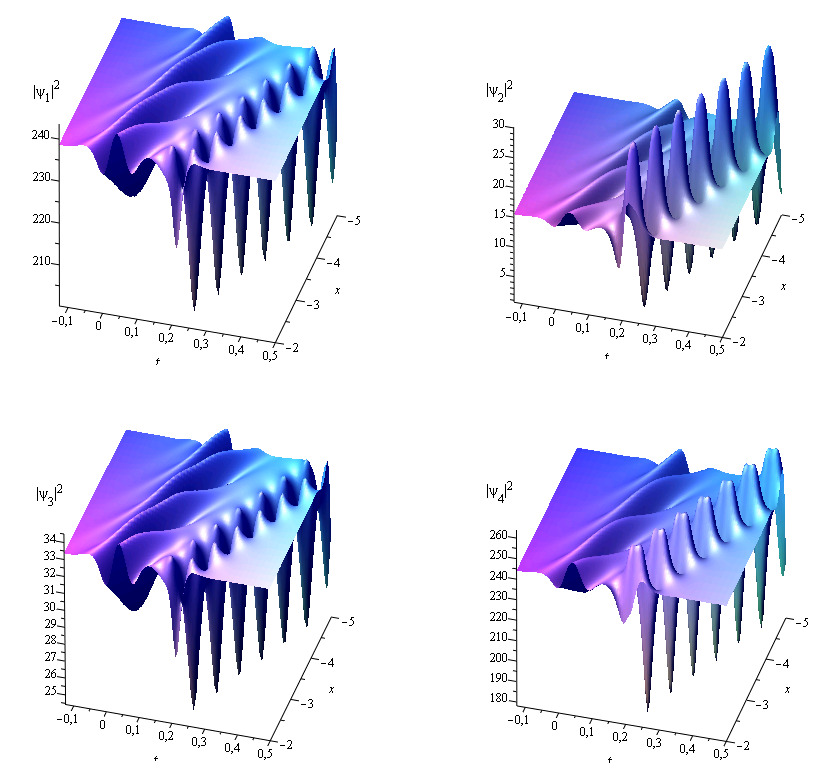}
\end{center}
\caption{2-breather of $4$-NLS$^{-++-}$. \label{2breather4NLS}}
\end{figure}

\begin{example} 
Figure \ref{breather4NLS} shows a breather solution of the 4-NLS$^{s}$ equation with $s=(-,-,+,-)$. It corresponds to the following choice of parameters:  $w_{a_{1}}=10,\,w_{a_{2}}=-5,\,w_{a_{3}}=-1/3,\,w_{a_{4}}=1/4,\,w_{a_{5}}=1/2,$ and $w_{u_{1}}\approx 0.55-0.11\mathrm{i}$ with  $f(w_{u_{1}})=2\mathrm{i}$, $w_{u_{2}}\approx-0.35+0.07\mathrm{i}$ with $f(w_{u_{2}})=-2\mathrm{i}$. 
\end{example}
\begin{example} 
Figure \ref{2breather4NLS} shows an elastic collision between two breather solutions of the 4-NLS$^{s}$ equation with $s=(-,+,+,-)$. It corresponds to the following choice of parameters: $w_{a_{1}}=1/3,\,w_{a_{2}}=3,\,w_{a_{3}}=1/7,\,w_{a_{4}}=2,\,w_{a_{5}}=1, \, w_{b_{1}}=-1,\,w_{b_{2}}=4,\,w_{b_{3}}=-2,\,w_{b_{4}}=0,$ and $w_{u_{1}}\approx 0.55-0.11\mathrm{i}$ with $f(w_{u_{1}})=2\mathrm{i}$, $w_{u_{2}}\approx-0.35+0.07\mathrm{i}$ with $f(w_{u_{2}})=-2\mathrm{i}$, $w_{u_{3}}\approx-0.91-0.52\mathrm{i}$ with $f(w_{u_{3}})=10-5\mathrm{i}$, and $w_{u_{4}}\approx14.46+5.32\mathrm{i}$ with $f(w_{u_{4}})=10+5\mathrm{i}$.
\end{example}

\subsubsection{$N$-rational breathers of n-NLS$^{s}$, for $1\leq N \leq 
n$.} 
Here we are interested in solutions of n-NLS$^{s}$ that can be expressed in the form of a ratio of two polynomials (modulo an exponential factor). These solutions, called rational breathers, are neither periodic in time nor in space, but are isolated in time and space. They are obtained from  breather solutions (\ref{breather n-NLS}) in the limit when the parameters $w_{v_{2k-1}}$ and $w_{u_{2k-1}}$ tend to each others,  as well as the parameters $w_{v_{2k}}$ and $w_{u_{2k}}$,  for $k=1,\ldots N$. An appropriate choice of the parameters $d_{i}$ in (\ref{breather n-NLS}) for $i=1,\ldots,2N$, leads to  limits of the form $0/0$ in the expression for the breather solutions. Thus, by performing a Taylor expansion of the numerator and denominator in (\ref{breather n-NLS}), one gets
a family of $N$-rational breather solutions of n-NLS$^{s}$.

\begin{proposition} 
Let $N,j\in\N$ satisfy $1\leq N \leq 
n$ and $1\leq j\leq n$. Let $f$ be a real meromorphic function (\ref{fct f}) of degree $n+1$ on the sphere, having $n+1$ real zeros $\{w_{a_{i}}\}_{i=1}^{n+1}$. Choose $\theta\in\R$ and take $\mathbf{\hat{d}}\in\C^{2N}$ such that $\overline{\hat{d}_{2k}}=\hat{d}_{2k-1}$ for $1\leq k\leq N$. Moreover, let $w_{u_{2k-1}},w_{v_{2k}}\in\C$, $1\leq k\leq N$, be complex conjugate critical points of the meromorphic function $f$, i.e.,  they are solutions of  $f'(w)=0$, which is equivalent to 
\begin{equation}
\sum_{i=1}^{n+1}\frac{1}{f'(w_{a_{i}})}\,\frac{1}{(w-w_{a_{i}})^{2}}=0.  \label{cond rat n-NLS}
\end{equation}
Put $s_{j}=\text{sign}(f'(w_{a_{n+1}})f'(w_{a_{j}}))$. Then the following functions give $N$-rational breathers of n-NLS$^{s}$
\begin{equation}
\psi_{j}(x,t)=A_{j}\,e^{\mathrm{i}\theta}\,\frac{\det(\mathbb{K}_{j,1})}{\det(\mathbb{K}_{j,0})}\,\exp\left\{ \mathrm{i}\,(-E_{j}\,x+F_{j}\,t)\right\}, \label{rat sol n-NLS}
\end{equation}
where $A_{j}=|q_{2}(a_{n+1},a_{j})|^{1/2}.$ For $\beta=0,1,$  $\mathbb{K}_{j,\beta}$ denotes a $2N\times2N$ matrix with entries $(\mathbb{K}_{j,\beta})_{i,k}$ given by:
\begin{eqnarray}
\text{- for $i$ and $k$ even:}& (\mathbb{K}_{j,\beta})_{ik}\hspace{-0.2cm}&=(1-\delta_{i,k})\,\frac{1}{w_{v_{i}}-w_{v_{k}}}-\delta_{i,k}\,(z_{k}+\beta\,\hat{r}_{j,k}) \nonumber\\
\text{- for $i$ even and $k$ odd:}& (\mathbb{K}_{j,\beta})_{ik}\hspace{-0.2cm}&=\frac{1}{w_{v_{i}}-w_{u_{k}}}\nonumber\\
\text{- for $i$ odd and $k$ even:}& (\mathbb{K}_{j,\beta})_{ik}\hspace{-0.2cm}&=-\,\frac{1}{w_{u_{i}}-w_{v_{k}}}
 \nonumber\\
\text{- for $i$ and $k$ odd:}& (\mathbb{K}_{j,\beta})_{ik}\hspace{-0.2cm}&=-\,(1-\delta_{i,k})\,\frac{1}{w_{u_{i}}-w_{u_{k}}}-\delta_{i,k}\,(z_{k}+\beta\,\hat{r}_{j,k}). \nonumber
\end{eqnarray}
Here $z_{k}$ is a linear function of the variables $x$ and $t$ given by
\[z_{k}=\mathrm{i}\,\hat{V}_{a_{n+1},k}\,x+\mathrm{i}\,\hat{W}_{a_{n+1},k}\,t-\hat{d}_{k} \]
for $k=1,\ldots,2N,$ where $\overline{ \hat{V}_{a_{n+1},2k}}= -\,\hat{V}_{a_{n+1},2k-1}$ and $\overline{ \hat{W}_{a_{n+1},2k}}= -\,\hat{W}_{a_{n+1},2k-1}$ with
\[\hat{V}_{a_{n+1},2k-1}=\frac{1}{f'(w_{a_{n+1}})}\,\frac{1}{(w_{a_{n+1}}-w_{u_{2k-1}})^{2}}, \quad \hat{W}_{a_{n+1},2k-1}=-\,\frac{1}{f'(w_{a_{n+1}})^{2}}\,\frac{2}{(w_{a_{n+1}}-w_{u_{2k-1}})^{3}}\]
for $k=1,\ldots,N$. Scalars $\hat{r}_{j,k}$ satisfy $\overline{ \hat{r}_{j,2k}}= -\,\hat{r}_{j,2k-1}$ and are given by
\begin{equation}
\hat{r}_{j,2k-1}=-\,\frac{w_{a_{n+1}}-w_{a_{j}}}{(w_{a_{n+1}}-w_{u_{2k-1}})\,(w_{a_{j}}-w_{u_{2k-1}})}\,. \nonumber
\end{equation}
Scalars $E_{j},F_{j}$ are defined by
\[E_{j}=\frac{1}{f'(w_{a_{n+1}})\,(w_{a_{j}}-w_{a_{n+1}})},\qquad F_{j}=-\,E_{j}^{2}+2\sum_{k=1}^{n}q_{2}(a_{n+1},a_{k}). \]
\label{prop rat sol n-NLS}
\end{proposition}

\begin{proof} To symplify the expression for the obtained solutions, apply the transformation (\ref{trans sol n-NLS}) to functions (\ref{breather n-NLS}) with $\beta$ and $\lambda$ as in (\ref{tranf breather}).
Let $\epsilon>0$ be a small parameter and define
$d_{k}=\epsilon \,\hat{d}_{k}+\mathrm{i}\pi,$
for $k=1,\ldots, 2N$. Moreover, assume
\begin{equation}
w_{v_{2k-1}}=w_{u_{2k-1}}+\epsilon \,\alpha_{v_{2k-1}}, \qquad w_{u_{2k}}=w_{v_{2k}}+\epsilon \,\alpha_{u_{2k}},  \label{limit sol rat n-NLS}
\end{equation}
for some $\alpha_{v_{2k-1}},\alpha_{u_{2k}}\in\C$, where $k=1,\ldots,N$.
Note that equation number $k$ of system (\ref{deg sum V}) can be written as
\begin{equation}
\sum_{j=1}^{n+1}\frac{1}{f'(w_{a_{j}})}\,\frac{f(w_{v_{k}})\,f(w_{u_{k}})}{(w_{a_{j}}-w_{v_{k}})\,(w_{a_{j}}-w_{u_{k}})}=-\,\frac{f(w_{v_{k}})-f(w_{u_{k}})}{w_{v_{k}}-w_{u_{k}}}\,. \label{k sum V}
\end{equation}
Hence, in the limit $\epsilon \rightarrow 0$, equation (\ref{k sum V}) becomes 
\[\sum_{j=1}^{n+1}\frac{1}{f'(w_{a_{j}})}\,\frac{f(w_{v_{2k-1}})^{2}}{(w_{a_{j}}-w_{v_{2k-1}})^{2}}=-\,f'(w_{v_{2k-1}}),\]
and
\[\sum_{j=1}^{n+1}\frac{1}{f'(w_{a_{j}})}\,\frac{f(w_{u_{2k}})^{2}}{(w_{a_{j}}-w_{u_{2k}})^{2}}=-\,f'(w_{u_{2k}}),\]
for $k=1,\ldots,N$.
Therefore, choose $w_{v_{2k-1}}$ and $w_{u_{2k}}$ to be distinct critical points of the meromorphic function $f$ for $k=1,\ldots N,$ i.e., they are solutions of $f'(w)=0$, in such way that equation (\ref{sum V}) holds in the limit considered here.
Since the condition $f'(w)=0$ is equivalent to solve a polynomial equation of degree $2n$, it follows that $1\leq N \leq n$.
Now take the limit $\epsilon \rightarrow 0$ in (\ref{breather n-NLS}). Note that parameters $\alpha_{v_{2k-1}},\alpha_{u_{2k}}$ cancel in this limit, and the degenerated functions take the form (\ref{rat sol n-NLS}). 
\end{proof}

\begin{remark}
\rm{Functions (\ref{rat sol n-NLS}) provide a family of rational breather solutions of n-NLS$^{s}$ depending on $N$ complex parameters $d_{k}$, a real parameter $\theta$, and a real meromorphic function $f$ (\ref{fct f}) defined by $2n+2$ real parameters, chosen such that $f$ admits complex conjugate critical points.}
\end{remark}

\begin{example} 
With the notation of Proposition \ref{prop rat sol n-NLS} the functions $\psi_{j}$ (\ref{rat sol n-NLS}) for $N=1$ are given by
\[\psi_{j}(x,t)=A_{j}\,e^{\mathrm{i}\theta}\,\frac{B+(z_{1}+\hat{r}_{j,1})(\overline{z_{1}}-\overline{\hat{r}_{j,1}})}{B+|z_{1}|^{2}}\,\exp\left\{ \mathrm{i}\,(-E_{j}x+N_{j}t)\right\},\]
where $B=\left(2\,\text{Im}(w_{u_{1}})\right)^{-2}.$
\end{example}
\begin{figure}[hbtp]  
\begin{center}
\includegraphics[width=120mm,height=95mm]{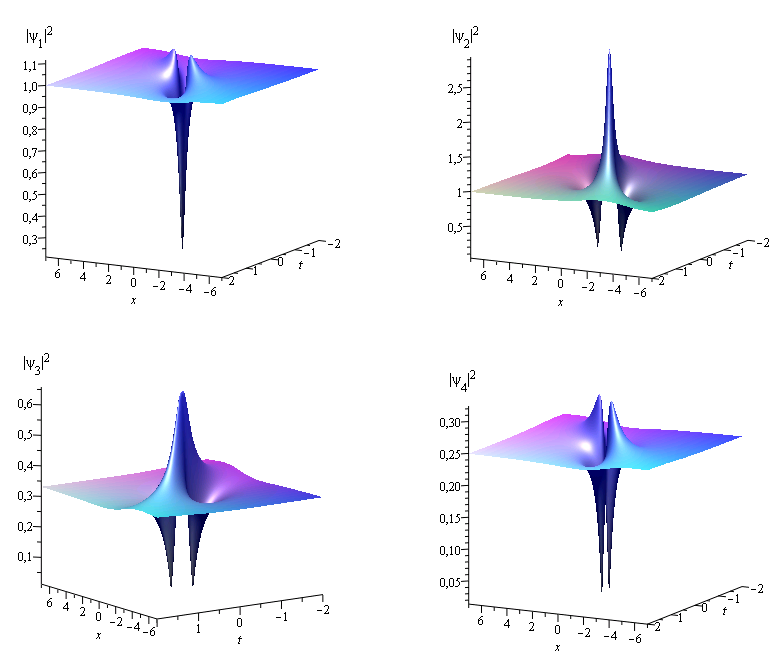}
\end{center}
\caption{Rational breather of $4$-NLS$^{++++}$. \label{rationalbreather4NLS}}
\end{figure}
\begin{figure}[hbtp]  
\begin{center}
\includegraphics[width=120mm,height=95mm]{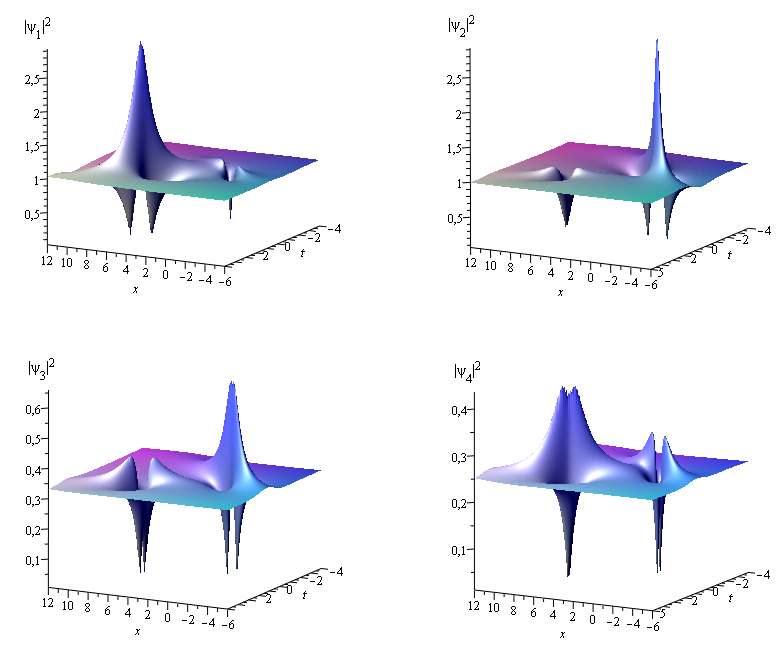}
\end{center}
\caption{2-rational breather of $4$-NLS$^{++++}$. \label{2rationalbreather4NLS}}
\end{figure}
\begin{example}
Figure \ref{rationalbreather4NLS} shows a rational breather solution of the 4-NLS$^{s}$ equation with $s=(+,+,+,+)$. It corresponds to the following choice of parameters: $k_{a_{k}}(w)=f'(w_{a_{k}})f(w)$ for $k=1,\ldots,n+1$, with $w_{a_{1}}=3,\,w_{a_{2}}=5,\,w_{a_{3}}=7,\,w_{a_{4}}=0,\,w_{a_{5}}=4,$ and $w_{u_{1}}\approx 4.53+0.56\mathrm{i}$ being a solution of $\sum_{i=1}^{n+1}(w-w_{a_{i}})^{-2}=0$. 
We observe that functions $\psi_{2}$ and $\psi_{3}$ coincide with the Peregrine breather well known in the scalar case \cite{Per}, whereas functions $\psi_{1},\psi_{4}$ belong to a new class of rational breathers which does not exist in the scalar case. This new type of rational breathers emerges due to the higher degree of the meromorphic function associated to the solutions of n-NLS$^{s}$ for $n>1$.
\end{example}
\begin{example}
Figure \ref{2rationalbreather4NLS} shows a 2-rational breather solution of the 4-NLS$^{s}$ equation with $s=(+,+,+,+)$. It corresponds to the following choice of parameters: $k_{a_{k}}(w)=f'(w_{a_{k}})f(w)$ for $k=1,\ldots,n+1$, with $w_{a_{1}}=3,\,w_{a_{2}}=5,\,w_{a_{3}}=7,\,w_{a_{4}}=0,\,w_{a_{5}}=4,$ and $w_{u_{1}}\approx4.53+0.56\mathrm{i}$, $w_{u_{3}}\approx 3.45+0.56\mathrm{i}$ being solutions of $\sum_{i=1}^{n+1}(w-w_{a_{i}})^{-2}=0$, and $d_{k}=10$. Variation of the parameters $d_{k}$ leads to a displacement in the $(x,t)$-plane of the rational breathers appearing in each of the pictures of Figure \ref{2rationalbreather4NLS}.
\end{example}

\section{Degenerate algebro-geometric solutions of the DS equations}

Solutions of the DS equations (\ref{DSintro})  in terms of elementary functions constructed here are obtained analogously to the solutions of the n-NLS equation, therefore some details will be omitted.
Let us introduce the function $\phi:=\Phi+\rho|\psi|^{2}$, where $\rho=\pm1$,  and the characteristic coordinates 
$\xi=\frac{1}{2}(x-\mathrm{i}\alpha \,y),$ $\eta=\frac{1}{2}(x+\mathrm{i}\alpha \,
y),$ $\alpha=\mathrm{i},1$ in (\ref{DSintro}).
In these coordinates the DS equations become
\begin{align}	
\mathrm{i}\,\psi_{t}+\frac{1}{2}(\partial_{\xi}^{2}+\partial_{\eta}^{2})\psi+2\,\phi\,\psi &=0,     \nonumber  \\
\partial_{\xi}\partial_{\eta}\phi+\frac{\rho}{2}(\partial_{\xi}^{2}+\partial_{\eta}^{2})|\psi|^{2} & =0.  \label{DS} 
\end{align}
Recall that DS1$^{\rho}$ denotes the case $\alpha=\mathrm{i}$ (here $\xi$ and $\eta$ are both 
real), and DS2$^{\rho}$ the case $\alpha=1$ (here  $\xi$ and $\eta$ are pairwise conjugate).

To construct  solutions of (\ref{DS})  in terms of elementary functions, let us first introduce its complexified version:
\begin{align}
	\mathrm{i}\,\psi_{t}+\frac{1}{2}(\psi_{\xi\xi}+\psi_{\eta\eta})+2\,\varphi\,\psi=0, \nonumber\\ \label{DS comp}
	-\mathrm{i}\,\psi^{*}_{t}+\frac{1}{2}(\psi^{*}_{\xi\xi}+\psi^{*}_{\eta\eta})+2\,\varphi\,\psi^{*}=0,\\ \nonumber
\varphi_{\xi\eta}+\frac{1}{2}((\psi\psi^{*})_{\xi\xi}+(\psi\psi^{*})_{\eta\eta})=0,
\end{align}
where $\varphi:=\Phi+\psi\psi^{*}$. This system reduces to (\ref{DS}) under the \textit{reality condition}: 
\begin{equation}
\psi^{*}=\rho\,\overline{\psi}, \label{real cond DS}
\end{equation}
which leads to $\varphi=\phi$. Theta-functional solutions of  (\ref{DS comp}) were studied in \cite{Kalla} and can be written in the following form.

\begin{theorem} Let $\mathcal{R}_{g}$ be a compact Riemann surface of genus $g>0$, and let $a,b\in\Rs_{g}$ be distinct points. Take arbitrary constants $\mathbf{D}\in\C^{g}$ and $A,\kappa_{1},\kappa_{2}\in\C\setminus\left\{0\right\}$, $h\in\C$. Denote by $\ell$ a contour connecting $a$ and $b$ which does not intersect cycles of the canonical homology basis. Then for any $\xi,\eta,t\in\C$, the following functions $\psi$, $\psi^{*}$ and $\varphi$ are solutions of system (\ref{DS comp})
\begin{align}
\psi(\xi,\eta,t)&=A\,\frac{\Theta(\mathbf{Z}-\mathbf{D}+\mathbf{r})}{\Theta(\mathbf{Z}-\mathbf{D})}\,\exp\left\{ -\mathrm{i}\left(G_{1}\,\xi+G_{2}\,\eta-G_{3}\,\tfrac{t}{2}\right)\right\},\nonumber\\
\psi^{*}(\xi,\eta,t)&=-\,\frac{\kappa_{1}\kappa_{2}\,q_{2}(a,b)}{A}\,\frac{\Theta(\mathbf{Z}-\mathbf{D}-\mathbf{r})}{\Theta(\mathbf{Z}-\mathbf{D})}\,\exp\left\{ \mathrm{i}\left(G_{1}\,\xi+G_{2}\,\eta-G_{3}\,\tfrac{t}{2}\right)\right\},\label{sol DS}\\
\varphi(\xi,\eta,t)&=\frac{1}{2}\,(\ln\Theta(\mathbf{Z-\mathbf{D}}))_{\xi\xi}+\frac{1}{2}\,(\ln\Theta(\mathbf{Z-\mathbf{D}}))_{\eta\eta}+\frac{h}{4}. \nonumber
\end{align}
Here $\mathbf{Z}=\mathrm{i}\,\kappa_{1}\mathbf{V}_{a}\,\xi-\mathrm{i}\,\kappa_{2}\mathbf{V}_{b}\,\eta+\mathrm{i}\,(\kappa^{2}_{1}\,\mathbf{W}_{a}-\kappa^{2}_{2}\,\mathbf{W}_{b})\,\frac{t}{2},$
where the vectors $\mathbf{V}_{a},\mathbf{V}_{b}$ and $\mathbf{W}_{a},\mathbf{W}_{b}$ were introduced in (\ref{exp hol diff}). Moreover $\mathbf{r}=\int_{\ell}\omega$, where $\omega$ is the vector of normalized holomorphic differentials, and the scalars $G_{1},G_{2},G_{3}$ are given by
\begin{equation}
G_{1}=\kappa_{1}\,K_{1}(a,b),\qquad G_{2}=\kappa_{2}\,K_{1}(b,a), \label{N12 DS}
\end{equation}
\begin{equation}
G_{3}=\kappa^{2}_{1}\,K_{2}(a,b)+\kappa^{2}_{2}\,K_{2}(b,a)+h. \label{N3 DS}
\end{equation}
Scalars $q_2(a,b), K_{1}(a,b),K_{2}(a,b)$ are defined in (\ref{q2}), (\ref{K1}), (\ref{K2}) respectively. 
\end{theorem}

\begin{remark}
\rm{In the case where vectors $\mathbf{V}_{a}$ and $\mathbf{V}_{b}$ satisfy $\mathbf{V}_{a}+\mathbf{V}_{b}=0$, as mentioned in \cite{Kalla}, solutions of the Davey-Stewartson equation become solutions of the NLS equation (\ref{NLS}) under an appropriate change of variables. }
\end{remark}

In this section, we study the behaviour of theta-functional solutions (\ref{sol DS}) of the complexified DS equations
when the Riemann surface 
degenerates into a Riemann surface of genus zero. Imposing the reality condition (\ref{real cond DS}), for particular choices of the parameters one gets well-known solutions such as multi-soliton, breather, rational breather, dromion and lump. This appears to be  the first time that such solutions of DS are derived from algebro-geometric solutions.

\subsection{Determinantal solutions of the complexified DS equations}

Here  solutions of the complexified system (\ref{DS comp}) are given as a quotient of two determinants. In the next subsections, this particular form will be more convenient to produce special solutions of the DS equations (\ref{DS}).

\begin{proposition} Let $k\in\N$ satisfy $1\leq k\leq g$. Let $w_{a},w_{b},w_{u_{k}},w_{v_{k}},h\in\C$, and $A,\kappa_{1},\kappa_{2}\in\C\setminus\left\{0\right\}$. Choose $\mathbf{d}\in\C^{g}$. Then the following functions are solutions of the system (\ref{DS comp}) 
\begin{align}    
\psi(\xi,\eta,t)&=A\,\frac{\det(\mathbb{T}_{1})}{\det(\mathbb{T}_{0})}\,\exp\left\{- \mathrm{i}\,(G_{1}\,\xi+G_{2}\,\eta-G_{3}\,\tfrac{t}{2})\right\}, \nonumber\\
\psi^{*}(\xi,\eta,t)&=-\,\frac{\kappa_{1}\,\kappa_{2}}{A\,(w_{a}-w_{b})^{2}}\,\frac{\det(\mathbb{T}_{-1})}{\det(\mathbb{T}_{0})}\,\exp\left\{ \mathrm{i}\,(G_{1}\,\xi+G_{2}\,\eta-G_{3}\,\tfrac{t}{2})\right\},  \label{deg sol DS}\\ 
\varphi(\xi,\eta,t)&=\frac{1}{2}\,(\ln\det(\mathbb{T}_{0}))_{\xi\xi}+\frac{1}{2}\,(\ln\det(\mathbb{T}_{0}))_{\eta\eta}+\frac{h}{4}.\nonumber
\end{align}
For $\beta=-1,0,1$, $\mathbb{T}_{\beta}$ denotes the $g\times g$ matrix with entries (\ref{element T}) with $z_{k}=Z_{k}-d_{k}+\beta\,r_{k}$. Here the scalars $r_{k}$ are given in (\ref{r}) and  
\begin{equation}
\mathbf{Z}=\mathrm{i}\,\kappa_{1}\mathbf{V}_{a}\,\xi-\mathrm{i}\,\kappa_{2}\mathbf{V}_{b}\,\eta+\mathrm{i}\,(\kappa^{2}_{1}\,\mathbf{W}_{a}-\kappa^{2}_{2}\,\mathbf{W}_{b})\,\frac{t}{2}  \label{deg Z DS}
\end{equation}
with
\begin{equation}
V_{c,k}=\frac{1}{w_{c}-w_{v_{k}}}-\frac{1}{w_{c}-w_{u_{k}}}, \qquad W_{c,k}=-\,\frac{1}{(w_{c}-w_{v_{k}})^{2}}+\frac{1}{(w_{c}-w_{u_{k}})^{2}},  \label{deg VW DS}
\end{equation}
where $c\in\{a,b\}$. 
The scalars $G_{1},G_{2},G_{3}$ are given by
\begin{equation}
G_{1}=\frac{\kappa_{1}}{w_{b}-w_{a}},\quad G_{2}=\frac{\kappa_{2}}{w_{a}-w_{b}}, \quad G_{3}=-\,G_{1}^{2}-G_{2}^{2}+h. \label{deg G DS}
\end{equation}
\end{proposition}

\begin{proof}
Consider solutions (\ref{sol DS}) of system (\ref{DS comp}) in the limit when the  Riemann surface degenerates to a Riemann surface of genus zero, as explained in Section 3. In this limit, choose the local parameters $k_{a}$ and $k_{b}$ near $a\in\Rs_{0}$ and $b\in\Rs_{0}$ to be the uniformization map between the degenerate Riemann surface $\Rs_{0}$ and the $w$-sphere. Hence, for any $w\in\Rs_{0}$ lying in a neighbourhood of $w_{a}\in\Rs_{0}$, $k_{a}(w)=w-w_{a}$. Therefore, quantities independent of variables $\xi,\eta$ and $t$ are obtained from (\ref{V})-(\ref{K2 deg}).
\end{proof}

\begin{remark}
\rm{Functions (\ref{deg sol DS}) give a family of  solutions of the complexified system, involving elementary functions only. These solutions depend on $3g+6$ complex parameters $w_{a},w_{b},h,A,\kappa_{1},\kappa_{2}$ and $w_{u_{k}},w_{v_{k}},d_{k}$. Varying these parameters we will obtain different types of physically interesting solutions investigated in the next subsections.}
\end{remark}

\subsection{Multi-solitonic solutions of the DS equations}

Soliton solutions of the DS equations were shown to be representable in terms of Wronskian determinants in \cite{AF}.
Single soliton and multi-soliton solutions  corresponding to the known one-dimensional solutions can be obtained from this representation. These solitons are pseudo-one-dimensional in the sense that in the $(x,y)$-plane, they have the same form as one-dimensional solitons in the
$(x,t)$-plane, but that they move with an angle with respect to the axes. The multi-soliton solution describes the interaction of many such solitons each propagating in different directions.
\\\\
In what follows $N\in\N$ with $N\geq 1$.

\subsubsection{Dark multi-soliton of DS1$^{\rho}$ and DS2$^{+}$} 

Here dark multi-solitons of the DS1$^{\rho}$ and DS2$^{+}$ equations are derived from functions (\ref{deg sol DS}) for an appropriate choice of the parameters.
They were investigated in \cite{YNSA}.
\\\\
Put $g=N$ and $A=|\kappa_{1}\kappa_{2}|^{1/2}\,|w_{a}-w_{b}|^{-1}$ in (\ref{deg sol DS}). Moreover, assume $h\in\R$ and $\mathbf{d}\in\R^{N}$.
\\\\
\textit{\textbf{Reality condition for DS1$^{\rho}$.}} Let us check that with the following choice of parameters, 
\begin{equation}
w_{a},w_{b}\in\R, \qquad \kappa_{1},\kappa_{2}\in\R\setminus\left\{0\right\}, \qquad
\overline{w_{v_{k}}}=w_{u_{k}}, \qquad k=1,\ldots, N, \label{H1 dark}
\end{equation}
functions $\psi$ and $\psi^{*}$ in (\ref{deg sol DS}) satisfy the reality condition $\psi^{*}=\rho\,\overline{\psi}$ with $\rho=-\,\text{sign}(\kappa_{1}\kappa_{2})$.
Indeed, this can be deduced from the fact that $G_{1},G_{2},G_{3}\in\R$, and
\begin{equation}
\overline{\det\left(\mathbb{T}_{\beta}\right)}=\det\left(\,\overline{\mathbb{T}_{\beta}}\,\right)=\det\left(\mathbb{T}_{-\beta}\right), \label{conj T}
\end{equation}
since $u$ and $v$ can be interchanged  in the proof of (\ref{T}). Therefore, functions $\psi$ and $\phi$ in (\ref{deg sol DS}) define dark multi-soliton solutions of DS1$^{\rho}$. 
\\\\
\textit{Smoothness.} The dark multi-soliton solutions obtained here are smooth because  
the denominator $\det(\mathbb{T}_{0})$ of functions $\psi$ and $\phi$ (\ref{deg sol DS}) consists of a finite sum of real exponentials (see (\ref{T})), since
$\xi,\eta,t$ are real. 

\begin{remark}
\rm{One gets a family of smooth dark multi-soliton of the DS1$^{\rho}$ equation, depending on $N+6$ real parameters  $w_{a},w_{b},h,\kappa_{1},\kappa_{2},d_{k}$, a phase $\theta$, and $N$ complex parameters $w_{u_{k}}$.}
\end{remark}

\noindent
\textit{\textbf{Reality condition for DS2$^{+}$.}} Let us check that with the following choice of parameters, 
\begin{equation}
\overline{w_{a}}=w_{b},\qquad \overline{\kappa_{1}}=\kappa_{2}, \qquad
w_{u_{k}}, w_{v_{k}}\in\R,\qquad k=1,\ldots, N,\label{H2 dark}
\end{equation}
the functions $\psi$ and $\psi^{*}$ (\ref{deg sol DS}) satisfy the reality condition $\psi^{*}=\overline{\psi}$.
With (\ref{H2 dark}), it is straightforward to see that (\ref{conj T}) is also satisfied. Moreover, since $\overline{G_{1}}=G_{2}$, $G_{3}\in\R$ and $(w_{a}-w_{b})^{2}<0$, the functions $\psi$ and $\psi^{*}$ (\ref{deg sol DS})
satisfy the reality condition $\psi^{*}=\overline{\psi}$. Therefore, they define dark multi-soliton solutions of DS2$^{+}$.
\\\\
\textit{Smoothness.} To get smooth solutions, additional conditions are needed 
to ensure that $\det(\mathbb{T}_{0})$ does not vanish for all complex conjugate
$\xi=\bar{\eta}$. For instance, if 
\[w_{v_{1}}< w_{u_{1}}< w_{v_{2}}< w_{u_{2}} < \ldots <w_{v_{N}}<w_{u_{N}},\]
the scalars $(\mathbb{B})_{ik}$ (\ref{B}) are real for any $i,k\in\left\{1,\ldots,N\right\}$. Therefore, the functions $\psi$ and $\phi$ (\ref{deg sol DS}) are smooth, since their denominator does not vanish as a finite sum of real exponentials.

\begin{remark}
\rm{One gets a family of smooth dark multi-soliton of the DS2$^{+}$ equation, depending on $3N+1$ real parameters  $h,w_{u_{k}},w_{v_{k}},d_{k}$, a phase $\theta$, and $2$ complex parameters $w_{a},\kappa_{1}$.}
\end{remark}


\subsubsection{Bright multi-soliton of DS1$^{\rho}$ and DS2$^{-}$} 

In this part we construct bright multi-soliton to the DS1$^{\rho}$ and DS2$^{-}$ equations.
It is well 
known that such solutions can be written in terms of a quotient of 
sums of exponentials, for which the modulus tends to zero if the 
spatial variables tend to infinity. 
\\\\
To get bright multi-soliton solutions, one degenerates once more solutions (\ref{deg sol DS}) of the complexified system. Put $g=2N$ and $A=|\kappa_{1}\kappa_{2}|^{1/2}\,|w_{a}-w_{b}|^{-1}$ in (\ref{deg sol DS}), and take $h\in\R$. 
\\\\
\textbf{\textit{Degeneration.}}
Choose a small parameter $\epsilon>0$ and define $d_{k}=-\ln \epsilon+\hat{d}_{k},$ for $k=1,\ldots,2N$, and 
\begin{equation}
\renewcommand{\arraystretch}{1.5}
\begin{array}{llll}
w_{u_{2k-1}}\hspace{-0.2cm}&=w_{a}+\epsilon\,\alpha_{u_{2k-1}}^{-1}\,(w_{a}-w_{b}),&\quad w_{v_{2k-1}}\hspace{-0.2cm}&=w_{b}+\epsilon\,\alpha_{v_{2k-1}}^{-1}\,(w_{a}-w_{b}),\\
w_{u_{2k}}\hspace{-0.2cm}&=w_{b}+\epsilon\,\alpha_{u_{2k}}^{-1}\,(w_{a}-w_{b}),& \quad w_{v_{2k}}\hspace{-0.2cm}&=w_{a}+\epsilon\,\alpha_{v_{2k}}^{-1}\,(w_{a}-w_{b}),
\end{array}\label{uv bright}
\end{equation}
for $k=1,\ldots,N$.
Moreover, put  $\kappa_{1}=\epsilon\,\hat{\kappa}_{1}\,(w_{a}-w_{b}),$ and $\kappa_{2}=\epsilon\,\hat{\kappa}_{2}\,(w_{a}-w_{b}).$
Consider in the determinant $\det(\mathbb{T}_{1})$ appearing in (\ref{deg sol DS}) the substitution
\[L_{2i}\longrightarrow L_{2i}-\frac{(\mathbb{T}_{1})_{2i,2}}{(\mathbb{T}_{1})_{2,2}}\,L_{2}\]
 for $i=2,\ldots,N$, where $L_{k}$ denotes the line number $k$ of the matrix $\mathbb{T}_{1}$ and $(\mathbb{T}_{1})_{i,k}$ the entries of this matrix. An analogous transformation has to be considered for the matrix $\mathbb{T}_{-1}$ appearing in function $\psi^{*}$. Now take the limit $\epsilon\rightarrow 0$ in (\ref{deg sol DS}). The function $\psi$ obtained in this limit has the form (\ref{bright DS}). Notice that in this limit, the dependence on the parameters $w_{a}$ and $w_{b}$ disappears.
\\\\
\textbf{\textit{Reality condition for DS1$^{\rho}$.}}
It is straightforward to see that, with the following choice of parameters, 
\begin{equation}
\hat{\kappa}_{1},\hat{\kappa}_{2}\in\R\setminus\left\{0\right\},\quad \overline{\hat{d}_{2k-1}}=\hat{d}_{2k}, \quad \overline{\alpha_{u_{2k-1}}}=\alpha_{v_{2k}}, \quad\overline{\alpha_{u_{2k}}}=\alpha_{v_{2k-1}},  \qquad k=1,\ldots, N,\label{H1}
\end{equation}
the functions $\psi$ and $\psi^{*}$ obtained in the limit considered here satisfy the reality condition $\psi^{*}=\rho\,\overline{\psi}$ with $\rho=-\,\text{sign}(\hat{\kappa}_{1}\hat{\kappa}_{2})$.
\\\\
\textbf{\textit{Reality condition for DS2$^{-}$.}}
In the same way, with the following choice of parameters, 
\begin{equation}
\overline{\hat{\kappa}_{1}}=\hat{\kappa}_{2}, \quad \overline{\hat{d}_{2k-1}}=\hat{d}_{2k},  \quad \overline{\alpha_{u_{2k-1}}}=\alpha_{u_{2k}}, \quad\overline{\alpha_{v_{2k-1}}}=\alpha_{v_{2k}},\qquad k=1,\ldots, N,\label{H2}
\end{equation}
the functions $\psi$ and $\psi^{*}$ obtained in the considered limit  satisfy the reality condition $\psi^{*}=-\,\overline{\psi}$.
\\\\
\textbf{\textit{The solutions.}}
Let $\theta\in\R$. 
With (\ref{H1}), the following functions of the variables $\xi,\eta,t$ obtained in the considered limit, give bright $N$-soliton solutions of DS1$^{\rho}$ where $\rho=-\,\text{sign}(\hat{\kappa}_{1}\hat{\kappa}_{2})$ and $\gamma=0$; because of (\ref{H2}) these functions define bright $N$-soliton solutions of DS2$^{-}$ where $\gamma=1$:
\begin{align}
\psi(\xi,\eta,t)&=\hat{A}\,e^{\mathrm{i}\theta}\,\frac{\det(\mathbb{K})}{\det(\mathbb{M})}, \nonumber\\
\phi(\xi,\eta,t)&=\frac{1}{2}\,(\ln\det(\mathbb{M}))_{\xi\xi}+\frac{1}{2}\,(\ln\det(\mathbb{M}))_{\eta\eta}+\frac{h}{4},  \label{bright DS}
\end{align}
where $\hat{A}=|\hat{\kappa}_{1}\hat{\kappa}_{2}|^{1/2}.$ Here $\mathbb{K}$ and $\mathbb{M}$ are $2N\times2N$ matrices with entries $(\mathbb{K})_{ik}$ and $(\mathbb{M})_{ik}$ given by:
\begin{eqnarray}
\text{- for $i$ and $k$ even:}& (\mathbb{K})_{ik}\hspace{-0.2cm}&=\delta_{i,k}-\delta_{2,i}\,\delta_{2,k}+\delta_{2,i}\,e^{\frac{1}{2}(z_{2}+z_{k}+\hat{r}_{2}+\hat{r}_{k})} \nonumber\\
&&\quad+\,\delta_{2,k}\,(\delta_{2,i}-1)\,e^{\frac{1}{2}(z_{i}-z_{2}+\hat{r}_{i}-\hat{r}_{2})}\nonumber\\ 
\text{- for $i$ even and $k$ odd:}& (\mathbb{K})_{ik}\hspace{-0.2cm}&=-\,\frac{\alpha_{u_{k}}^{2}}{\alpha_{v_{i}}-\alpha_{u_{k}}}\frac{\alpha_{v_{2}}-\alpha_{v_{i}}}{\alpha_{v_{2}}-\alpha_{u_{k}}}\,e^{\frac{1}{2}(z_{i}+z_{k}+\hat{r}_{i}+\hat{r}_{k})}\nonumber\\
\text{- for $i$ odd and $k$ even:}& (\mathbb{K})_{ik}\hspace{-0.2cm}&=-\,\frac{\alpha_{v_{i}}\,\alpha_{u_{k}}}{\alpha_{v_{i}}-\alpha_{u_{k}}}\,e^{\frac{1}{2}(z_{i}+z_{k}+\hat{r}_{i}+\hat{r}_{k})}
 \nonumber\\
\text{- for $i$ and $k$ odd:}& (\mathbb{K})_{ik}\hspace{-0.2cm}&=\delta_{i,k}, \nonumber\\\nonumber\\
\text{- for $i,k$ even, or $i,k$ odd:}&(\mathbb{M})_{ik}\hspace{-0.2cm}&=\delta_{i,k}\nonumber\\ 
\text{- otherwise:}&(\mathbb{M})_{ik}\hspace{-0.2cm}&=(-1)^{i+1}\,\frac{\alpha_{v_{i}}\,\alpha_{u_{k}}}{\alpha_{v_{i}}-\alpha_{u_{k}}}\,e^{\frac{1}{2}(z_{i}+z_{k})}.
 \nonumber
\end{eqnarray}
Here $z_{k}$ is a linear function of the variables $\xi,\eta$ and $t$ given by
\[z_{2k-1}=\mathrm{i}\,\hat{\kappa}_{1}\,\alpha_{u_{2k-1}}\,\xi+\mathrm{i}\,\hat{\kappa}_{2}\,\alpha_{v_{2k-1}}\,\eta+\mathrm{i}\left(\hat{\kappa}_{1}^{2}\,\alpha_{u_{2k-1}}^{2}+\hat{\kappa}_{2}^{2}\,\alpha_{v_{2k-1}}^{2}\right)\frac{t}{2}-\hat{d}_{2k-1}+\gamma\,\frac{\mathrm{i}\pi}{2},\]
\[z_{2k}=-\,\mathrm{i}\,\hat{\kappa}_{1}\,\alpha_{v_{2k}}\,\xi-\mathrm{i}\,\hat{\kappa}_{2}\,\alpha_{u_{2k}}\,\eta-\mathrm{i}\left(\hat{\kappa}_{1}^{2}\,\alpha_{v_{2k}}^{2}+\hat{\kappa}_{2}^{2}\,\alpha_{u_{2k}}^{2}\right)\frac{t}{2}-\hat{d}_{2k}+\gamma\,\frac{\mathrm{i}\pi}{2},\]
for $k=1,\ldots,N$. Moreover, the scalars $\hat{r}_{k}$ are defined by
\[\hat{r}_{k}=(-1)^{k}\ln\left\{-\alpha_{v_{k}}\alpha_{u_{k}}\right\}, \qquad k=1,\ldots,2N.\]

\begin{remark}
\rm{i) With (\ref{H1}), functions (\ref{bright DS}) give a family of bright multi-soliton solutions of the DS1$^{\rho}$ equation depending on  $3N$ complex parameters $\hat{d}_{2k-1},\alpha_{u_{2k-1}}, \alpha_{u_{2k}}$ and $4$ real parameters $h,\theta,\hat{\kappa}_{1},\hat{\kappa}_{2}$. 
\\
ii) With (\ref{H2}), functions (\ref{bright DS}) provide a family of bright multi-soliton solutions of the DS2$^{-}$ equation depending on  $3N+1$ complex parameters $\hat{d}_{2k-1},\alpha_{u_{2k-1}}, \alpha_{v_{2k-1}},\hat{\kappa}_{1}$ and $2$ real parameters $h,\theta$.}
\end{remark}


\subsection{Breather and rational breather solutions of the DS equations}

The breather solutions of the DS equation were found in \cite{TA}. Here a family of breather solutions and rational breather solutions
of the DS1 equation are derived from algebro-geometric solutions. These solutions resemble their $1+1$ dimensional analogues. In particular,
the profiles of the corresponding solutions of the DS equation in the $(x,y,t)$ coordinates look as those in the $(x,t)$ coordinates extended along a spatial variable $y$.

\subsubsection{Multi-Breathers of DS1$^{\rho}$} 

The $N$-breather solution obtained here corresponds to an elastic interaction between $N$ breathers. Put $g=2N$ and  $A=|\kappa_{1}\kappa_{2}|^{1/2}\,|w_{a}-w_{b}|^{-1}$ in (\ref{deg sol DS}). 
It is straightforward to see that with the following choice of parameters, 
\begin{equation}
 w_{a},w_{b},h\in\R, \quad \kappa_{1},\kappa_{2}\in\R\setminus\left\{0\right\}, \quad \overline{d_{2k-1}}=d_{2k}, \quad \overline{w_{v_{2k}}}=w_{u_{2k-1}}, \quad \overline{w_{v_{2k-1}}}=w_{u_{2k}},  \label{uv breather DS}
\end{equation}
for $k=1,\ldots,N $, functions $\psi$ and $\psi^{*}$ (\ref{deg sol DS}) satisfy the reality condition $\psi^{*}=\rho\,\overline{\psi}$ with $\rho=-\,\text{sign}(\kappa_{1}\kappa_{2})$. Therefore, analogously to the n-NLS equation, functions $\psi$ and $\phi$ in (\ref{deg sol DS})
give $N$-breather solutions of DS1$^{\rho}$.

\begin{remark}
\rm{One gets a family of breather solutions of DS1$^{\rho}$ depending on $3N$ complex parameters $d_{2k-1},$
$w_{u_{2k-1}},w_{u_{2k}}$ and $6$ real parameters $w_{a},w_{b},h,\kappa_{1},\kappa_{2}$ and a phase $\theta$.}
\end{remark}

\begin{figure}[hbtp]  
\begin{center}
\includegraphics[height=65mm]{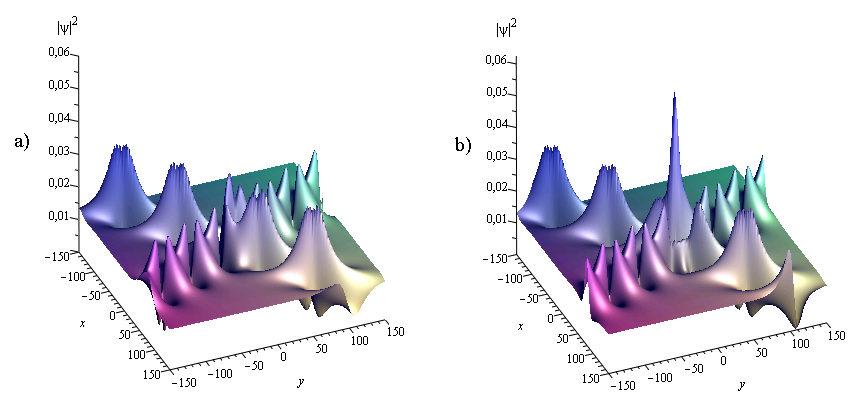}
\end{center}
\caption{$2$-breather of DS1$^{-}$ at a) $t=0$, b) $t=45$. \label{2breatherDS1}}
\end{figure}

\begin{example}
Figure \ref{2breatherDS1} shows the evolution in time of $2$-breather solution of DS1$^{-}$ with the following choice of parameters: $w_{a}=8,\, w_{b}=-1,\, w_{u_{1}}=5-2\mathrm{i},\,w_{u_{2}}=2+\mathrm{i},\,w_{u_{3}}=3-\mathrm{i},\,w_{u_{4}}=1+4\mathrm{i},\,\kappa_{1}=\kappa_{2}=1,\,d_{k}=h=0$.
\end{example}

\subsubsection{Multi-rational breathers of DS1$^{\rho}$} 

In this part, we deal with rational solutions (modulo an exponential factor) of the DS1$^{\rho}$ 
equation. These solutions are
obtained as limiting cases of the breather solutions. The $N$-rational solutions describe elastic interaction between $N$ rational breathers, and are expressed as a quotient of two polynomials of degree $N$ in the 
variables $\xi,\eta,t$. 
\\\\
Assume $g=2N$ and put $A=|\kappa_{1}\kappa_{2}|^{1/2}\,|w_{a}-w_{b}|^{-1}$ in (\ref{deg sol DS}).
\\\\
\textit{\textbf{Degeneration.}}
Let $\epsilon>0$ be a small parameter and define  $d_{k}=\epsilon \,\hat{d}_{k}+\mathrm{i}\pi,$ for $k=1,\ldots, 2N,$ 
and 
\begin{equation}
w_{v_{2k-1}}=w_{u_{2k-1}}+\epsilon \,\alpha_{v_{2k-1}}, \qquad w_{u_{2k}}=w_{v_{2k}}+\epsilon \,\alpha_{u_{2k}}  \label{limit sol rat}
\end{equation}
for $k=1,\ldots,N$.
It is straightforward to see that $\det(\mathbb{T}_{\beta})\approx \epsilon^{2N}P_{\beta}$, where $P_{\beta}$ is a polynomial of degree $2N$ with respect to the variables $\xi,\eta$ and $t$.
Now take the limit $\epsilon \rightarrow 0$ in (\ref{deg sol DS}). 
The function $\psi$ obtained in this limit is an $N$-rational breather solution of DS1$^{\rho}$ given by (\ref{rat sol DS}).
\\\\
\textit{\textbf{Reality condition.}}
Imposing the following constraints on the parameters:
\begin{equation}
w_{a},w_{b},h\in\R,\quad \kappa_{1},\kappa_{2}\in\R\setminus\left\{0\right\},\quad \overline{\hat{d}_{2k}}=\hat{d}_{2k-1}, \quad \overline{w_{u_{2k-1}}}=w_{v_{2k}}, \quad k=1,\ldots, N, \label{cond real sol rat}
\end{equation}
it can be seen that the functions $\psi$ and $\psi^{*}$ (\ref{deg sol DS}) in the considered limit 
satisfy the reality condition $\psi^{*}=\rho \,\overline{\psi}$, with $\rho=-\,\text{sign}(\kappa_{1}\kappa_{2})$.
\\\\
\textit{\textbf{The solutions.}}
Let $\theta\in\R$. Then the following degenerated functions define $N$-rational breather solutions of DS1$^{\rho}$ 
\begin{align}
\psi(\xi,\eta,t)&=A\,e^{\mathrm{i}\theta}\,\frac{\det(\mathbb{K}_{1})}{\det(\mathbb{K}_{0})}\,\exp\left\{-\mathrm{i}\,(G_{1}\,\xi+G_{2}\,\eta-G_{3}\,\tfrac{t}{2})\right\}, \nonumber\\
\phi(\xi,\eta,t)&=\frac{1}{2}\,(\ln\det(\mathbb{K}_{0}))_{\xi\xi}+\frac{1}{2}\,(\ln\det(\mathbb{K}_{0}))_{\eta\eta}+\frac{h}{4},\label{rat sol DS}
\end{align}
where $\mathbb{K}_{\beta}$, with $\beta=0,1,$ is a $2N\times2N$ matrix with entries $(\mathbb{K}_{\beta})_{ik}$ given by
\begin{eqnarray}
\text{- for $i$ and $k$ even:}& (\mathbb{K}_{\beta})_{ik}\hspace{-0.2cm}&=(1-\delta_{i,k})\,\frac{1}{w_{v_{i}}-w_{v_{k}}}-\delta_{i,k}\,(z_{k}+\beta\,\hat{r}_{k})\nonumber\\ 
\text{- for $i$ even and $k$ odd:}& (\mathbb{K}_{\beta})_{ik}\hspace{-0.2cm}&=\frac{1}{w_{v_{i}}-w_{u_{k}}}\nonumber\\
\text{- for $i$ odd and $k$ even:}& (\mathbb{K}_{\beta})_{ik}\hspace{-0.2cm}&=-\,\frac{1}{w_{u_{i}}-w_{v_{k}}}
 \nonumber\\
\text{- for $i$ and $k$ odd}:& (\mathbb{K}_{\beta})_{ik}\hspace{-0.2cm}&=-\,(1-\delta_{i,k})\,\frac{1}{w_{u_{i}}-w_{u_{k}}}-\delta_{i,k}\,(z_{k}+\beta\,\hat{r}_{k}). \nonumber
\end{eqnarray}
Here $z_{k}$ is a linear function of the variables $\xi,\eta$ and $t$ given by 
\[z_{k}=\mathrm{i}\,\kappa_{1}\hat{V}_{a,k}\,\xi-\mathrm{i}\,\kappa_{2}\hat{V}_{b,k}\,\eta+\mathrm{i}\left(\kappa_{1}^{2}\,\hat{W}_{a,k}-\kappa_{2}^{2}\,\hat{W}_{b,k}\right)\frac{t}{2}-\hat{d}_{k}. \]
Moreover, for $c\in\left\{a,b\right\}$, the scalars $\hat{V}_{c,k},\hat{W}_{c,k}$ and $\hat{r}_{k}$ satisfy
$\overline{\hat{V}_{c,2k}}=\hat{V}_{c,2k-1}$, $\overline{\hat{W}_{c,2k}}=\hat{W}_{c,2k-1}$ and $\overline{\hat{r}_{2k}}=\hat{r}_{2k-1}$, and are given by:
\[\hat{V}_{c,2k-1}=\frac{1}{(w_{c}-w_{u_{2k-1}})^{2}},\qquad  \hat{W}_{c,2k-1}=-\,\frac{2}{(w_{c}-w_{u_{2k-1}})^{3}}, \] \[\hat{r}_{2k-1}=-\,\frac{w_{a}-w_{b}}{(w_{a}-w_{u_{2k-1}})\,(w_{b}-w_{u_{2k-1}})},\]
for $k=1,\ldots,N$.
Constants $G_{1},G_{2},G_{3}$ are given in (\ref{deg G DS}).

\begin{remark}
\rm{Functions (\ref{rat sol DS}) give a family of rational solutions of DS1$^{\rho}$ depending on $2N$ complex parameters $d_{2k-1},w_{u_{2k-1}}$ and $6$ real parameters $w_{a},w_{b},h,\theta,\kappa_{1},\kappa_{2}$.}
\end{remark}

\begin{figure}[hbtp]  
\begin{center}
\includegraphics[height=45mm]{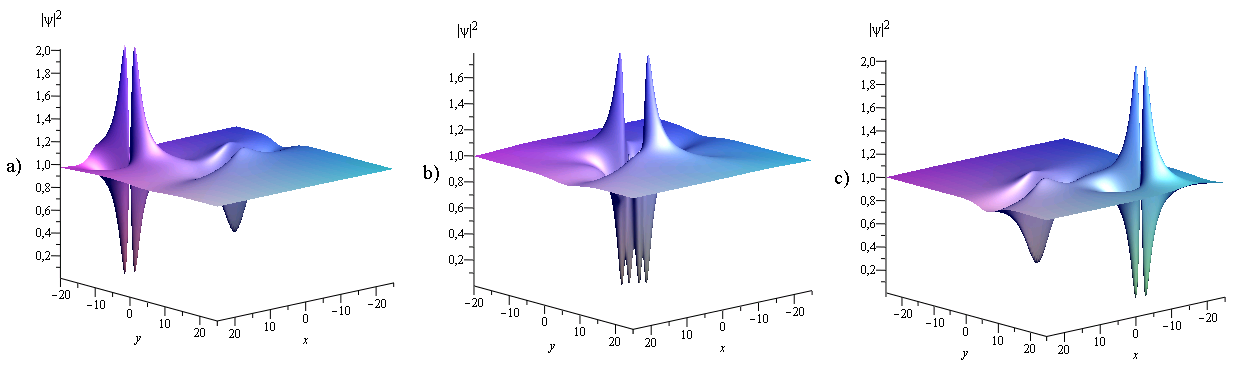}
\end{center}
\caption{$2$-rational breather of DS1$^{-}$ at a) $t=-5$, b) $t=0$, c) $t=5$. \label{2rationalbreatherDS1}}
\end{figure}

\begin{example} Figure \ref{2rationalbreatherDS1} shows the evolution in time of the $2$-rational breather solution of DS1$^{-}$ with the following choice of parameters: $w_{a}=2,\, w_{b}=1,\, w_{u_{1}}=2\mathrm{i},\,w_{u_{3}}=2+\mathrm{i},\,\kappa_{1}=\kappa_{2}=1,\,d_{k}=h=0$.
\end{example}

\begin{figure}[hbtp]  
\begin{center}
\includegraphics[height=85mm]{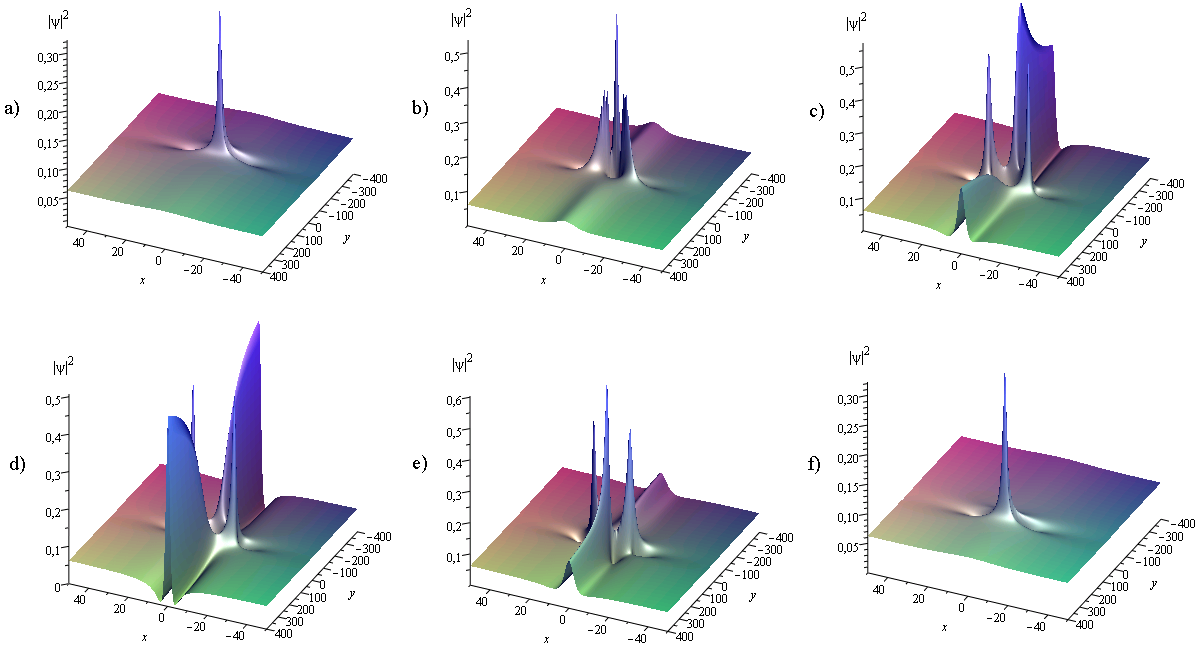}
\end{center}
\caption{Interaction between a line rational breather and a rational breather of DS1$^{-}$ at a) $t=-50$, b) $t=-20$, c) $t=-5$, d) $t=0$, e) $t=10$, f) $t=50$. The rational breather propagates in the same direction as the line breather. \label{2rationalbreatherDS1bis}}
\end{figure}

\begin{figure}[hbtp]  
\begin{center}
\includegraphics[height=85mm]{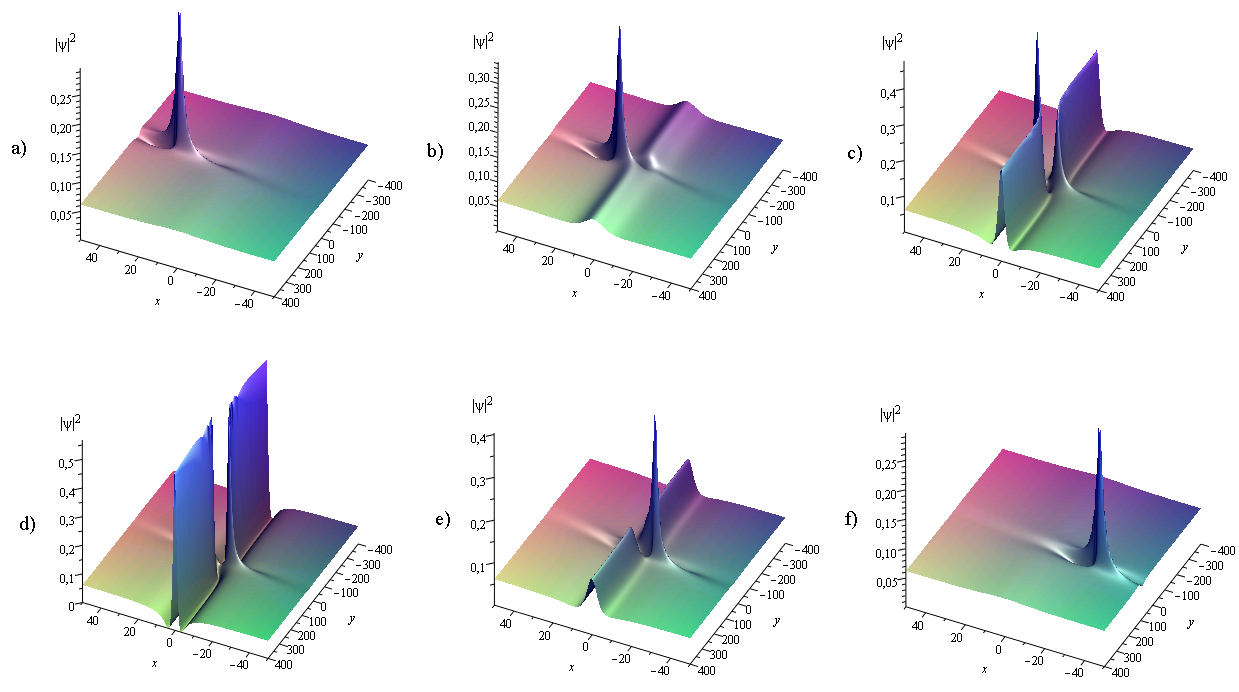}
\end{center}
\caption{Interaction between a line rational breather and a rational breather of DS1$^{-}$ at a) $t=-50$, b) $t=-20$, c) $t=-5$, d) $t=0$, e) $t=10$, f) $t=50$. The rational breather propagates transversally to the direction of the line breather. \label{2rationalbreatherDS1bisbis}}
\end{figure}

\begin{example} Figure \ref{2rationalbreatherDS1bis} (resp. Figure \ref{2rationalbreatherDS1bisbis}) shows the interaction between a line rational breather and a rational breather solution of DS1$^{-}$ with the following choice of parameters: $w_{a}=2,\, w_{b}=-2,\, w_{u_{1}}=3\mathrm{i}$ (resp. $w_{u_{1}}=3\mathrm{i}+1$) , $w_{u_{3}}=2\mathrm{i},\,\kappa_{1}=\kappa_{2}=1,\,d_{k}=h=0$. By \textit{line rational breather} we denote a growing and decaying mode localized only in one direction.
\end{example}

\subsection{Dromion and lump solutions of the DS equations}

Here we construct the dromion solution of DS1$^{\rho}$ and the lump solution of DS2$^{-}$ which correspond to  solutions 
localized in all directions of the plane.
These solutions arise by suitable degenerations of solutions (\ref{deg sol DS}) to the complexified system, and by imposing the reality condition $\overline{\psi^{*}}=\rho\,\psi$. This appears to be the first time that such  solutions are obtained as limiting cases of theta-functional solutions.

\subsubsection{Dromion of DS1$^{\rho}$} 

Boiti et al. \cite{BLMP} have shown that the DS1 equation has solutions that decay exponentially in all directions. 
The solutions they obtained can move along any direction in the plane, and the only effect of their interactions is a shift in their position, independently of their relative initial position in the plane.
Later, Fokas and Santini \cite{FS, Sant} pointed out that by an appropriate choice of the boundary conditions, the localized solitons (called  ''dromions'') of the DS1 equation possess properties which are different from the properties of one-dimensional solitons, namely, the performed 
solutions  do not preserve their form upon interaction. For a particular choice of their spectral parameters, they recovered solutions previously derived by Boiti et al. For details on the theory of dromion solutions the reader is referred to \cite{RL} and references therein. In this section
we explore how the simplest dromion solution can be derived from algebro-geometric solutions.
\\\\
Let us consider solutions of the complexified system obtained in (\ref{deg sol DS}). Assume $g=4$ and put $A=|\kappa_{1}\kappa_{2}|^{1/2}\,|w_{a}-w_{b}|^{-1}.$
\\\\
\textit{\textbf{Degeneration.}}
Choose a small parameter $\epsilon>0$ and define $d_{k}=-\ln(\epsilon)+ \hat{d}_{k}$ for $k=1,\ldots, 4,$ and
\begin{align}
w_{u_{1}}&=\epsilon\, \alpha_{u_{1}}, &\quad w_{u_{2}}&=w_{a}+\epsilon\, \alpha_{u_{2}}, &\quad w_{u_{3}}&=w_{b}+\epsilon \,\alpha_{u_{3}},&\quad w_{u_{4}}&=\epsilon\, \alpha_{u_{4}},\nonumber\\
w_{v_{1}}&=w_{a}+\epsilon\, \alpha_{v_{1}},&\quad w_{v_{2}}&=\epsilon\, \alpha_{v_{2}}, &\quad w_{v_{3}}&=\epsilon\, \alpha_{v_{3}},&\quad
w_{v_{4}}&=w_{b}+\epsilon\, \alpha_{v_{4}}.
 \label{deg var dromion}
\end{align}
Moreover, put $\kappa_{1}=\epsilon\,\hat{\kappa}_{1}\,\alpha_{v_{1}}$ and $\kappa_{2}=\epsilon\,\hat{\kappa}_{2}\,\alpha_{u_{3}}$. 
Now consider the limit $\epsilon \rightarrow 0$ in (\ref{deg sol DS}). The functions $\psi$ and $\phi$ obtained in this limit are given by  (\ref{dromion DS1}).
\\\\
\textit{\textbf{Reality condition.}}
Choose $w_{a},w_{b},h,\theta\in\R$ and $\hat{\kappa}_{1},\hat{\kappa}_{2}\in\R\setminus\left\{0\right\}$. Moreover, assume 
\begin{equation}
 \overline{\hat{d}_{2k}}=\hat{d}_{2k-1},  \quad
\overline{\alpha_{v_{2k-1}}}=\alpha_{u_{2k}}, \quad \overline{\alpha_{v_{2k}}}=\alpha_{u_{2k-1}}, \quad k=1,2. \label{var dromion}
\end{equation} 
Put $\rho=-\,\text{sign}(\hat{\kappa}_{1}\hat{\kappa}_{2})$. With (\ref{var dromion}), it can be seen that the degenerated functions $\psi$ and $\psi^{*}$  obtained in the considered limit satisfy the reality condition  $\psi^{*}=\rho\,\overline{\psi}$. Therefore, the following degenerated functions give the dromion solution of DS1$^{\rho}$ 
\begin{align}
\psi(\xi,\eta,t)&=\hat{A}\,e^{\mathrm{i}\theta}\,\frac{e^{z_{1}+z_{3}}}{\varphi(\xi,\eta,t)}, \nonumber\\
\phi(\xi,\eta,t)&=\frac{1}{2}\,\partial_{\xi\xi}\ln\left\{\varphi(\xi,\eta,t)\right\}+\frac{1}{2}\,\partial_{\eta\eta}\ln\left\{\varphi(\xi,\eta,t)\right\}+\frac{h}{4},\label{dromion DS1}
\end{align}
where 
\[\varphi(\xi,\eta,t)=1+A_{1}\,e^{2\,\text{Re}(z_{1})}+A_{2}\,e^{2\,\text{Re}(z_{3})}+A_{3}\,e^{2\,\text{Re}(z_{1})+2\,\text{Re}(z_{3})}.\]
Here $z_{k}$ is a linear function of the variables $\xi,\eta,t$ given by
\begin{equation}
z_{1}=-\,\mathrm{i}\,\frac{\hat{\kappa}_{1}}{\alpha_{v_{1}}}\,\xi-\mathrm{i}\,\frac{\hat{\kappa}_{1}^{2}}{\alpha_{v_{1}}^{2}}\,\frac{t}{2}-\hat{d}_{1},\qquad
z_{3}=-\,\mathrm{i}\,\frac{\hat{\kappa}_{2}}{\alpha_{u_{3}}}\,\eta-\mathrm{i}\,\frac{\hat{\kappa}_{2}^{2}}{\alpha_{u_{3}}^{2}}\,\frac{t}{2}-\hat{d}_{3}.\nonumber
\end{equation}
Constants $\hat{A}, A_{1},A_{2}$ and $A_{3}$ are given by
\[\hat{A}=|\hat{\kappa}_{1}\hat{\kappa}_{2}|^{1/2}\,\frac{w_{a}\,w_{b}}{(\alpha_{v_{3}}-\alpha_{u_{1}})\,\alpha_{v_{1}}\alpha_{u_{3}}},\qquad A_{1}=\frac{w_{a}}{4\,\text{Im}(\alpha_{v_{1}})\,\text{Im}(\alpha_{u_{1}})},\]\[ A_{2}=\frac{w_{b}}{4\,\text{Im}(\alpha_{v_{3}})\,\text{Im}(\alpha_{u_{3}})},
\qquad
A_{3}=A_{1}\,A_{2}+\frac{w_{a}\,w_{b}}{4\,\text{Im}(\alpha_{v_{1}})\,\text{Im}(\alpha_{u_{3}})}
\,\frac{1}{|\alpha_{u_{1}}-\alpha_{v_{3}}|^{2}}.\]
\\
Moreover, in the case where $A_{1}>0,\,A_{2}>0$ and $A_{3}>0$, functions (\ref{dromion DS1}) are smooth solutions of DS1$^{\rho}$.

\begin{remark}
\rm{i) Functions (\ref{dromion DS1}) define a family of dromion solutions of DS1$^{\rho}$ depending on $6$ complex parameters 
$\hat{d}_{1},\hat{d}_{3},\alpha_{u_{1}},\alpha_{v_{1}},\alpha_{u_{3}},\alpha_{v_{3}}$ and $6$ real parameters $w_{a},w_{b},\hat{\kappa}_{1},\hat{\kappa}_{2},h,\theta$. 
\\
ii) In the case where $\alpha_{u_{1}},\alpha_{v_{3}}\in\R$, one gets localized breathers, namely, the solution oscillates with respect to the time variable (modulus of $\psi$ is constant with respect to $t$).}
\end{remark}

%
%

Different degenerations can be investigated for larger values of $g$. The performed functions lead to particular solutions such as dromions which move along sets of straight and curved trajectories, as well as oscillating dromion solutions. We do not discuss these solutions here.

\subsubsection{Lump of DS2$^{-}$}
The lump solutions were discovered in \cite{MZBIM} for the KP1 equation, and have been extensively studied. Arkadiev et al. \cite{APP} have constructed a family of travelling
waves (the lump solutions) of DS2$^{-}$ that we rediscover here.
\\

Let us consider functions $\psi,\psi^{*},\phi$ given in (\ref{deg sol DS}), assume $g=2$ and put $A=|\kappa_{1}\kappa_{2}|^{1/2}|w_{a}-w_{b}|^{-1}.$ Moreover, consider the following transformation which leaves the system (\ref{DS comp}) invariant:
\begin{align}
\psi(\xi,\eta,t)&\rightarrow\psi\left(\xi+\beta_{1}\,t,\eta+\beta_{2}\,t,t\right)\,\exp\left\{ -\mathrm{i}\left(\beta_{1}\,\xi+\beta_{2}\,\eta+\left(\beta_{1}^{2}+\beta_{2}^{2}\right)\tfrac{t}{2}\right)\right\}, \nonumber
\\
\psi^{*}(\xi,\eta,t)&\rightarrow\psi^{*}\left(\xi+\beta_{1}\,t,\eta+\beta_{2}\,t,t\right)\,\exp\left\{ \mathrm{i}\left(\beta_{1}\,\xi+\beta_{2}\,\eta+\left(\beta_{1}^{2}+\beta_{2}^{2}\right)\tfrac{t}{2}\right)\right\}, \nonumber 
\\
\phi(\xi,\eta,t)&\rightarrow\phi\left(\xi+\beta_{1}\,t,\eta+\beta_{2}\,t,t\right), \label{trans sol DS bis}
\end{align}
where $\beta_{i}=\mu_{i}\,\kappa_{i}^{-1}$ for some $\mu_{i}\in\C$.
\\\\
\textbf{\textit{Degeneration.}} 
Choose a small parameter $\epsilon>0$ and define $d_{k}= \mathrm{i}\pi+\epsilon\,\hat{d}_{k},$ for $k=1,2,$
and
\begin{align}
w_{v_{1}}=w_{a}+\epsilon \,\alpha_{v_{1}},& \qquad w_{u_{1}}=w_{a}+\epsilon\, \alpha_{u_{1}}, \nonumber\\
w_{v_{2}}=w_{b}+\epsilon \,\alpha_{v_{2}}, &\qquad w_{u_{2}}=w_{b}+\epsilon\, \alpha_{u_{2}}. \nonumber
\end{align}
Moreover, put $\kappa_{k}=\epsilon^{2}\,\hat{\kappa}_{k},$ and
$\mu_{k}=\epsilon^{2}\,\hat{\mu}_{k}$ for $k=1,2.$ Now take the limit $\epsilon \rightarrow 0$ in (\ref{trans sol DS bis}). The functions $\psi$ and $\phi$  obtained in this limit are given in (\ref{lump sol bis}).
\\\\
\textbf{\textit{Reality condition.}} 
Choose $w_{a},w_{b}\in\C$ such that $\overline{w_{a}}=-\,w_{b}$ or $w_{a},w_{b}\in\R$. Take $h,\theta\in\R$ and assume
\begin{equation}
\overline{\hat{\kappa}_{1}}=\hat{\kappa}_{2}, \quad \overline{\hat{\mu}_{1}}=\hat{\mu}_{2},\quad \overline{\hat{d}_{1}}=\hat{d}_{2}, \quad \overline{\alpha_{v_{1}}}=\alpha_{v_{2}}, \quad \overline{\alpha_{u_{1}}}=\alpha_{u_{2}}.\nonumber
\end{equation}
With this choice of parameters, it can be seen that the functions $\psi$ and $\psi^{*}$ obtained in the limit considered here satisfy the reality condition $\psi^{*}=-\,\overline{\psi}$. 
\\\\
\textbf{\textit{The solutions.}} 
Therefore, the following degenerated functions provide smooth solutions of DS2$^{-}$ 
\begin{align}
\psi(x,y,t)&=\frac{\hat{A}\,e^{\mathrm{i}\theta}}{\hat{B}+|z_{1}|^{2}}\,\exp\left\{-  \mathrm{i}\left(2\,\text{Re}(\beta_{1}\,\xi)+\text{Re}(\beta_{1}^{2})\,t\right)\right\}, \nonumber\\
\phi(x,y,t)&=\frac{1}{2}\,\partial_{\xi\xi}\ln\left\{\hat{B}+|z_{1}|^{2}\right\}+\frac{1}{2}\,\partial_{\overline{\xi}\overline{\xi}}\ln\left\{\hat{B}+|z_{1}|^{2}\right\}+\frac{h}{4},\label{lump sol bis}
\end{align}
where $\xi=x+\mathrm{i}y$ and $\beta_{1}=\hat{\mu}_{1}\,\hat{\kappa}_{1}^{-1}$. Here  $z_{1}=\mathrm{i}\,\hat{V}_{a,1}\left(\hat{\kappa}_{1}\,\xi+\hat{\mu}_{1}\,t\right)-\hat{d}_{1} $ and
\[\hat{V}_{a,1}=-\,\frac{\alpha_{u_{1}}-\alpha_{v_{1}}}{\alpha_{u_{1}}\alpha_{v_{1}}},\qquad
\hat{A}= \frac{|\hat{\kappa}_{1}|\,|\alpha_{u_{1}}-\alpha_{v_{1}}|^{2}}{|w_{a}-w_{b}|\,\alpha_{v_{1}}\,\overline{\alpha_{u_{1}}}}, \qquad \hat{B}=\frac{|\alpha_{u_{1}}-\alpha_{v_{1}}|^{2}}{(w_{b}-w_{a})^{2}}\,.\]
\\
\textbf{\textit{Simplifications.}}
To simplify (\ref{lump sol bis}), put 
\[\hat{d}_{1}=-\,\frac{\mathrm{i}\,\mu}{\hat{V}_{a,1}\,\hat{\kappa}_{1}}, \quad \nu=\frac{\alpha_{u_{1}}\overline{\alpha_{v_{1}}}}{|\hat{\kappa}_{1}|\,|w_{a}-w_{b}|},\quad \lambda=\beta_{1},\]
for arbitrary $\mu\in\C$.
In this way, functions (\ref{lump sol bis}) become
\begin{align}
\psi(x,y,t)&=\nu\,\frac{\exp\{-2\mathrm{i}\,\text{Re}(\lambda \,\xi)-\mathrm{i}\,\text{Re}(\lambda^{2})\,t+\mathrm{i}\theta\}}{|\xi+\lambda\,t+\mu|^{2}+|\nu|^{2}}, \nonumber\\
\phi(x,y,t)&=\frac{1}{2}\,\partial_{\xi\xi}\ln\left\{|\xi+\lambda\,t+\mu|^{2}+|\nu|^{2}\right\}+\frac{1}{2}\,\partial_{\overline{\xi}\overline{\xi}}\ln\left\{|\xi+\lambda\,t+\mu|^{2}+|\nu|^{2}\right\}+\frac{h}{4},\label{lump DS}
\end{align}
where $\xi=x+\mathrm{i}y$. Here $\lambda,\nu,\mu$ are arbitrary complex constants, and $\theta,h\in\R$. 
Solutions (\ref{lump DS}) coincide with the lump solution previously obtained in \cite{APP}.

%
%
%
%

\section{Outlook}
In this paper, various classes of  solutions to the multi-component NLS equation and the DS equations  in terms of elementary functions have been presented as limiting cases of algebro-geometric solutions discussed in a previous paper \cite{Kalla}.
We did not construct all families of solutions present in the literature, but we believe that different degenerations will lead to interesting new or  known solutions that are not presented here. 

In particular, future investigations might address bright multi-solitons of n-NLS with inelastic collision.   
This novel type of inelastic collision, which is not observed in $1+1$ dimensional soliton systems, follows from a family of bright soliton solutions having more parameters than the ones presented here with standard elastic collision. We believe that also this kind of solutions arises from algebro-geometric solutions after suitable degenerations.

\vspace{1cm}
\thanks{I thank C. Klein who interested me in the subject, and V.~Shramchenko for carefully reading the
manuscript and providing valuable hints.  I am grateful to D.~Korotkin and V. Matveev for useful discussions and hints. 
This work has been supported in part by the project FroM-PDE funded by the European
Research Council through the Advanced Investigator Grant Scheme, the Conseil R\'egional de Bourgogne
via a FABER grant and the ANR via the program ANR-09-BLAN-0117-01. }


\begin{thebibliography}{99}


\bibitem{APT} M.J. Ablowitz, B. Prinari, A.D. Trubatch, \textit{Integrable Nonlinear Schrödinger Systems and their Soliton
Dynamics}, Dynamics of PDE Vol.\textbf{1}, No.3, 239--299 (2004).
\bibitem{AS} M.J. Ablowitz, H. Segur, \textit{Solitons and the Inverse Scattering Transform}, SIAM, Philadelphia, PA (1981).
\bibitem{Ak et Al} N.N. Akhmediev, V.M. Eleonskii,  N.E. Kulagin, \textit{First-order exact solutions
of the nonlinear Schrödinger equation}, Teoret. Mat. Fiz. \textbf{72}, 2:183--196 (1987). English
translation: Theoret. Math. Phys. \textbf{72}, 2:809--818 (1987).
\bibitem{AKG} Andonowati, N. Karjanto, E. van Groesen, \textit{Extreme wave phenomena in
down-stream running modulated waves}, Appl. Math. Model. \textbf{31}, 1425--1443 (2007).
\bibitem{AF} D. Anker, N.C. Freeman, \textit{ On the soliton solutions of the Davey-Stewartson equation for
long waves}, Proc. R. Soc. London vol. A \textbf{360}, 529--540 (1978).
\bibitem{APP} V.A. Arkadiev, A.K. Pogrebkov, M.C. Polivanov, \textit{Inverse scattering transform and soliton
solutions for Davey-Stewartson II equation}, Physica D \textbf{36},  189--197 (1989).
\bibitem{BBEIM} E. Belokolos, A. Bobenko, V. Enolskii, A. Its, V. Matveev, \textit{Algebro-geometric approach to nonlinear integrable equations}, Springer Series in nonlinear dynamics (1994).
\bibitem{BLMP} M. Boiti, J. Leon, L. Martina, F. Pempineili, \textit{Scattering of localized solitons in the plane}, Phys. Lett. A \textbf{132},  432--439 (1988).
\bibitem{Chen} D.Y. Chen,\textit{ Introduction to Solitons}, Science Press, Beijing (2006).
\bibitem{Des} A. Degasperis, \textit{Solitons}, Am. J. Phys. \textbf{66}, 486--497 (1998).
\bibitem{DM} P. Dubard, P. Gaillard, C. Klein, V.B. Matveev, \textit{On multi-rogue wave solutions of the NLS equation and positon solutions of the KdV equation}, Eur. Phys. J. Special Topics Vol. \textbf{185}, 247--258 (2010).
\bibitem{DS} A. Davey, K. Stewartson, {\it On three-dimensional packets of surface waves}, Proc. R. Soc. Lond. A {\bf 388}, 101--110 (1974).
\bibitem{EKK} V. Eleonskii, I. Krichever, N. Kulagin, \textit{Rational multisoliton solutions
to the NLS equation}, Soviet Doklady 1986 sect. Math. Phys. V. \textbf{287},
606--610 (1986).
\bibitem{Fay} J. Fay, {\it Theta functions on Riemann surfaces}, Lecture Notes in Mathematics \textbf{352} (1973).
\bibitem{FS} A.S. Fokas, P.M. Santini, \textit{Dromions and a boundary value problem for the Davey-Stewartson 1 equation}, Physica D \textbf{44},
 99 (1990).
\bibitem{F} N.C. Freeman, \textit{Soliton Solutions of Non-linear Evolution Equations}, IMA J. Appl. Math. \textbf{32}, 125 (1984).
\bibitem{FN} N.C. Freeman, J.J.C. Nimmo,\textit{ A method of obtaining the N-soliton solution of the Boussinesq equation in terms of a wronskian}, Phys. Lett. A \textbf{95}, 1  (1983).
\bibitem{GGMK} C.S. Gardner, J.M. Greene, R. Miura, M. Kruskal, \textit{Method for Solving the Korteweg-de Vries Equation}, Comm. Appl. Math. \textbf{27}, 97 (1974). 
\bibitem{HPD} K.L. Henderson, D.H. Peregrine, J.W. Dold, \textit{Unsteady water wave modulations:
fully nonlinear solutions and comparison with the nonlinear Schrödinger equation},
Wave Motion \textbf{29}, 341--361 (1999).
\bibitem{H} R. Hirota, \textit{Exact Solution of the Korteweg-de Vries Equation for Multiple Collisions of solitons}, Phys. Rev. Lett. \textbf{27}, 1192 (1971).
\bibitem{IRS} A.R. Its, A.V. Rybin, M.A. Salle, \textit{Exact integration of nonlinear
Schrödinger equation}, Teore. i Mat. Fiz. V. \textbf{74}, N. 1, 29--45 (1988).
\bibitem{Kalla} C. Kalla, \textit{New degeneration of Fay's identity and its application to integrable systems}, arXiv:1104.2568v1 [math-ph] (April 2011).
\bibitem{KLDA} T. Kanna, M. Lakshmanan, P. Tchofo Dinda, N. Akhmediev, \textit{Soliton collisions with shape change by intensity redistribution in mixed coupled nonlinear Schrödinger equations}, Phys. Rev. E \textbf{73} (2006).
\bibitem{Ma} Y.C. Ma, \textit{The perturbed plane-wave solutions of the cubic Schrödinger equation}, Stud.
Appl. Math. \textbf{60}, 1:43--58 (1979).
\bibitem{Man} S.V. Manakov, \textit{On the theory of two-dimensional stationary self-focusing of electromagnetic waves}, Sov. Phys. JETP \textbf{38}, 248 (1974).
\bibitem{MZBIM} S.V. Manakov, V.E. Zakharov, L.A. Bordag, A.R. Its,
 V.B. Matveev, \textit{Two Dimensional Solitons of the Kadomtsev-Petviashvili Equation and Their Interaction}, Phys. Lett. A \textbf{63}, 205 (1977).
\bibitem{Mat} Y. Matsuno, {\it Multiperiodic and multisoliton solutions of a nonlocal nonlinear 
Schrödinger equation for envelope waves}, Phys. Lett. A \textbf{278}, 53 (2000).
\bibitem{Matv} V.B. Matveev, M.A. Salli, \textit{Darboux Transformations and Solitons},
Springer Series in Nonlinear Dynamics, Springer-Verlag, Berlin (1991).
\bibitem{Mum} D. Mumford, {\it Tata Lectures on Theta. I and II.}, Progress in Mathematics, 28 and 43,
respectively. Birkhäuser Boston, Inc., Boston, MA, (1983 and 1984).
\bibitem{NF} J.J.C. Nimmo, N.C. Freeman, \textit{The use of Bäcklund transformations in obtaining N-soliton solutions in Wronskian form}, J. Phys. A: Math. Gen. \textbf{17}, 1415 (1984).
\bibitem{NMPZ} S.~Novikov, S.~Manakov, L.~Pitaevskii, 	V.~Zakharov, \textit{Theory of Solitons - The Inverse	Scattering Method}, Consultants Bureau: New York (1984).
\bibitem{OOS} A.R. Osborne, M. Onorato, M. Serio, \textit{The nonlinear dynamics of rogue waves
and holes in deep-water gravity wave trains}, Phys. Lett. A \textbf{275}, 386--393 (2000).
\bibitem{Per} D.H. Peregrine, \textit{Water waves, nonlinear Schrödinger equations and their solutions},
J. Austral. Math. Soc. Ser. B \textbf{25}, 1:16--43 (1983).
\bibitem{PZ} A.D. Polyanin, V.F. Zaitsev, \textit{Handbook of Nonlinear Partial Differential Equations}, Chapman and
Hall/CRC, Boca Raton (2004).
\bibitem{RL2} R. Radhakrishnan, M. Lakshmanan, \textit{Bright and dark soliton solutions to coupled nonlinear Schrödinger equations}, J. Phys. A, Math. Gen. \textbf{28}, 2683--2692 (1995).
\bibitem{RL3} R. Radhakrishnan, M. Lakshmanan, \textit{Inelastic Collision and Switching of
Coupled Bright Solitons in Optical Fibers}, Phys. Rev. E \textbf{56}, 2213 (1997).
\bibitem{RL} R. Radhakrishnan, M. Lakshmanan, \textit{Localized Coherent Structures and Integrability in a Generalized
$(2 + 1)$-Dimensional Nonlinear Schrödinger Equation}, Chaos, Solitons and Fractals \textbf{8},  p. 17 (1997).
\bibitem{RSL} R. Radhakrishnan, R. Sahadevan, M. Lakshmanan, {\it Integrability and singularity structure of coupled nonlinear Schrödinger 
equations}, Chaos, Solitons and Fractals \textbf{5}, No. 12,  2315--2327 (1995).
\bibitem{Sant} P.M. Santini, \textit{Energy exchange of interacting coherent structures in multidimensions},
Physica D \textbf{41}:26--54 (1990).
\bibitem{TA} M. Tajiri, T. Arai, \textit{Periodic soliton solutions to the Davey-Stewartson equation},
Proc. Inst. Math. Natl. Acad. Sci. Ukr. \textbf{30}, 1:210--217 (2000).
\bibitem{YNSA} N. Yoshida, K. Nishinari, J. Satsuma,  K. Abe, \textit{A new type of soliton behavior of the Davey-Stewartson equations in a plasma system,} J. Phys. A \textbf{31}, 3325 (1998).
\bibitem{ZSh} V. Zakharov, A. Shabat, {\it Exact theory of two-dimensional self-focusing and one-dimensional self-modulation of
waves in nonlinear media}, Soy. Phys. JETP \textbf{34}, 62--69 (1972).



\end{thebibliography}
\end{document}